\documentclass{article}

\usepackage{arxiv}

\usepackage[utf8]{inputenc}
\usepackage[T1]{fontenc}
\usepackage{microtype}
\usepackage{url}
\usepackage{amsmath}
\usepackage{amssymb}
\usepackage{amsfonts}
\usepackage{bm}
\usepackage{nicefrac}
\usepackage{graphicx}
\usepackage{float}
\usepackage{subcaption}
\usepackage{booktabs}
\usepackage{dcolumn}
\usepackage{xcolor}
\usepackage[numbers,square,sort&compress]{natbib}
\usepackage{hyperref}
\usepackage{doi}

\title{A Neural Implicit Flow Super-Resolution Approach for Reacting Flows}

\author{
Dakshaka Vishwajith Amarathunga \\
\textnormal{Department of Mechanical and Aerospace Engineering} \\
\textnormal{North Carolina State University} \\
\textnormal{Campus Box 7910, Raleigh, NC 27695, USA}
\And
Lanre M. Oshinowo \\
\textnormal{Saudi Arabian Oil Co., Res. \& Dev. Ctr.} \\
\textnormal{P.O. Box 62, Dhahran 31311, Saudi Arabia}
\AND
Ji-Woong Park \\
\textnormal{Aramco Americas, Aramco Research Center} \\
\textnormal{Detroit, Novi, MI 46535, USA}
\And
Tarek Echekki\thanks{Corresponding author:
\texttt{techekk@ncsu.edu}} \\
\textnormal{Department of Mechanical and Aerospace Engineering} \\
\textnormal{North Carolina State University} \\
\textnormal{Campus Box 7910, Raleigh, NC 27695, USA}
}

\renewcommand{\shorttitle}{Neural Implicit Flow Super-Resolution}

\date{}

\hypersetup{
    pdftitle={A Neural Implicit Flow Super-Resolution Approach for Reacting Flows},
    pdfauthor={Dakshaka Vishwajith Amarathunga,
               Lanre M. Oshinowo,
               Ji-Woong Park,
               Tarek Echekki},
    pdfsubject={Neural Implicit Flow Super-Resolution for Reacting Flows},
    pdfkeywords={Neural Implicit Flow, Super-Resolution,
                 Reacting Flows, Machine Learning, Combustion}
}

\begin{document}
\maketitle

\begin{abstract}
We present a novel framework for reconstructing fine-resolution scalar fields from coarse-resolution data in 2D unsteady flames using an extension of neural implicit flow (NIF). The framework leverages the novel architecture of NIF to develop an efficient representation of scalar fields in spatio-temporal reacting flows. Different strategies are adopted within the proposed framework to establish super-resolution for multiscalar fields in combustion. They include 1) a reconfiguration of the input and output fields, 2) the use of latent composition space based on nonlinear principal component analysis (NLPCA) to reconstruct the fine-resolution fields, and the conversion from nonlinear principal components (NLPCs) to species and temperature contours, 3) and the adoption of multiple ParameterNets in the NIF architecture, each accommodating one of the NLPCs. The framework was validated on a 2D unsteady Bunsen-stabilized lean premixed methane-air flame with forcing at the inlet. The results, using a coarsening ratio of 8 in a given direction, show that the adoption of multiple ParameterNets yields more accurate super-resolution than a single ParameterNet without any additional cost in the hyperparameters.
\end{abstract}

\maketitle

\section{Introduction}
\label{sec:introduction}
Turbulent combustion flows span a wide range of length and time scales. Large-eddy simulations (LES) emerged as a strategy to overcome the limitations of capturing these scales. They involve spatial filtering of the length scales, where large scales are resolved through the solution of the filtered governing equations for flow, species, and energy transport, while the contributions of the subgrid scales are modeled. 

In turbulent combustion flows, important physics resides in the unresolved scales, rendering closure even more challenging than the case of non-reacting flows. Many concepts that have evolved from RANS approaches have been extended to the LES of combustion flows. Alternatively, several offline or {\it in situ} data-driven approaches have been proposed to capture the unresolved physics from LES-resolved quantities in simulations.

In addition to LES, other coarse-grained strategies that do not require spatial filtering have been proposed~\cite{BarSinai2019, Kochkov2021, Zhuang2021}. The methods target the predictions of accurate and temporally resolved solutions on a coarse grid, and inherently require an account of the contribution of unresolved scales. These approaches lend themselves to data-driven strategies where prior fine-resolution data is used to construct mappings from coarse to fine-resolution fields.

Many of the data-driven deep learning techniques that have been implemented for the reconstruction of fine-resolution information from coarse-resolution information (e.g., LES) in turbulent flows could be extended to turbulent combustion flows. These include the implementation of variants of convolutional neural network networks~\cite{LeCun1998, Fukami2019} and generative adversarial neural networks (GANs)~\cite{Goodfellow2020} as effective super-resolution tools for turbulent flows where fine-grained information is extracted from coarse-grid information. 

In turbulent combustion flows, earlier work by Nikolaou et al.~\cite{Nikolaou2018} for the progress variable field, Lapeyre et al.~\cite{Lapeyre2019} for the flame surface density, and Seltz et al.~\cite{Seltz2019} for the source and scalar flux terms, focused on implementing a CNN for constructing an SGS closure.  Variants of GANs, including super-resolution GANs (SRGANs), enhanced SRGANS (ESRGANs)~\cite{Wang-ESRGAN} and physics-informed  ESRGANs (PIESRGANs), have been used as data-driven methods to reconstruct SGS scalar and velocity information from coarse LES solutions. These include the studies by Bode et al.~\cite{Bode2020}, Grenga et al.~\cite{Grenga2023}, and Nista et al.~\cite{Nista2024, Nista2025}. More recently, frameworks based on graph neural networks (GNNs) have been proposed for super-resolution by Barwey et al.\cite{Barwey2025} and Dash~\cite{Dash2026}.

CNN- and GAN-based techniques possess complementary features that make them suitable for reconstructing fine-grained information from coarse-grained fields in turbulent reacting and non-reacting flows, such as SGS information in LES. CNNs are well suited for learning local spatial features and mapping these features across different resolutions. In GAN-based super-resolution approaches, the generator network produces reconstructed high-resolution fields, while the discriminator network attempts to distinguish reconstructed fields from reference high-resolution data. This adversarial training, often combined with pointwise or physics-informed reconstruction losses, can improve the realism and sharpness of reconstructed flow structures. CNNs also often serve as foundational components within GAN architectures.

Recently, Pan et al.~\cite{Pan2023} proposed neural implicit flow (NIF). NIF is a mesh-agnostic machine learning framework that was developed specifically for spatio-temporal flows. Accordingly, the network structure decouples the spatial characteristics of flows from time and additional parameters. This decoupling is implemented using two fully connected neural-network subnetworks: a ShapeNet and a ParameterNet. 

The performance of the NIF model is demonstrated on the Rayleigh–Taylor instability by Pan et a.~\cite{Pan2023}, indicating the NIF capability for nonlinear dimensionality reduction on adaptive mesh data. The sparse pointwise sensor measurements from the adaptive mesh are utilized as an input to the ParameterNet, and the model reconstructs the entire flow field from the coordinates provided as an input to the ShapeNet. This setup compressed the temporal solutions into a one-dimensional latent space representation and achieved a superior accuracy over the Single Value Decomposition (SVD) and the Convolutional AutoEncoder (CAE).  

In contrast, in the present study, the NIF model is utilized for super-resolution of turbulent reacting flows, which is fundamentally different from dimensionality reduction. In the proposed framework, the model learns a mapping between the coarse-grid field (e.g., temperature, species mass fraction) and the fine-grid field to resolve the small-scale structures. The ParameterNet or ParameterNets utilize the coarse-grid field and generate(s) the parameters for the ShapeNet, which reconstructs the fine-resolution field based on a dense spatial grid. This framework leverages NIF's mesh-agnostic ability to capture the multi-physics of combustion while recovering SGS features filtered out in the coarse-grid solution.

The proposed approach is designed to efficiently construct a finer resolution of the coarse fields using only the spatial coordinates, thus requiring much less data. We propose the use of a modified NIF architecture~\cite{Pan2023} to construct higher resolution data. The architecture is designed to implement a cost-effective tool for reconstructing unresolved physics while exploiting the NIF features designed for the spatio-temporal flows. In the present study, the NIF-based super-resolution framework is demonstrated in 2D using unsteady simulations of lean methane-air premixed flames in a Bunsen burner configuration. Flow forcing at the inlet results in flame structures and features (including curvature, unsteady strain rates and flame-flame interactions) that are typical in turbulent flames. The paper is organized as follows. In the following section, Sec.~\ref{sec:methods}, we introduce and motivate the choice of the modified NIF architecture. In Sec.~\ref{sec:results}, we present the model problem of a turbulent burner-stabilized premixed flame and present results based on {\it a priori} comparisons of reconstructed fine-resolution scalar fields from coarse-resolution data based only on one scalar.

%\subsection{Tables}
%Tables, however small, must be numbered sequentially in the order in which they are mentioned in the text. Words \textit {table~1, table~2} should be lower case throughout. See table~\ref{tab:kd} for an example.
%\begin{table}
%  \begin{center}
%\def~{\hphantom{0}}
%  \begin{tabular}{lccc}
%      $a/d$  & $M=4$   &   $M=8$ & Callan \etal \\[3pt]
%       0.1   & 1.56905 & ~~1.56~ & 1.56904\\
%       0.3   & 1.50484 & ~~1.504 & 1.50484\\
%       0.55  & 1.39128 & ~~1.391 & 1.39131\\
%       0.7   & 1.32281 & ~10.322 & 1.32288\\
%       0.913 & 1.34479 & 100.351 & 1.35185\\
%  \end{tabular}
%  \caption{Values of $kd$ at which trapped modes occur when $\rho(\theta)=a$.}
%  \label{tab:kd}
%  \end{center}
%\end{table}

\section{Methodology}\label{sec:methods}
This section presents the methodological framework for the proposed NIF-based super-resolution approach for reacting flows. The framework consists of two main components. First, an extended Neural Implicit Flow (NIF) architecture is developed to enable efficient reconstruction of multiple scalar fields through a structured latent representation. This representation is achieved using nonlinear principal component analysis (NLPCA) {\it via} the implementation of an autoencoder architecture. The encoder component corresponds to the NLPCA, while the decoder part reproduces the original vector of thermochemical scalars, which corresponds to temperature and species mass fractions. Second, a subdomain-based learning strategy is introduced to capture localized flow features and promote domain-independent learning. These components collectively define the core modeling approach used for reconstructing high-resolution thermochemical fields from coarse-resolution inputs. By focusing on the local scalar and flow structures, this learning strategy generalizes the training model to a broad range of conditions, such as different inlet geometries and flow conditions or different flow conditions (e.g., turbulence intensity). Third, and given the multi-component nature of combustion flows and the inherent correlation between thermochemical scalars, we adopt multiple ParameterNets instead of the original single ParameterNet implementation of NIF. Each ParameterNet handles input from a single nonlinear principal component (NLPC) as outlined below.
\subsection{Extended NIF framework}\label{subsec:nif}
\begin{figure}[t]
        \centering
        \includegraphics[width=0.65\textwidth]{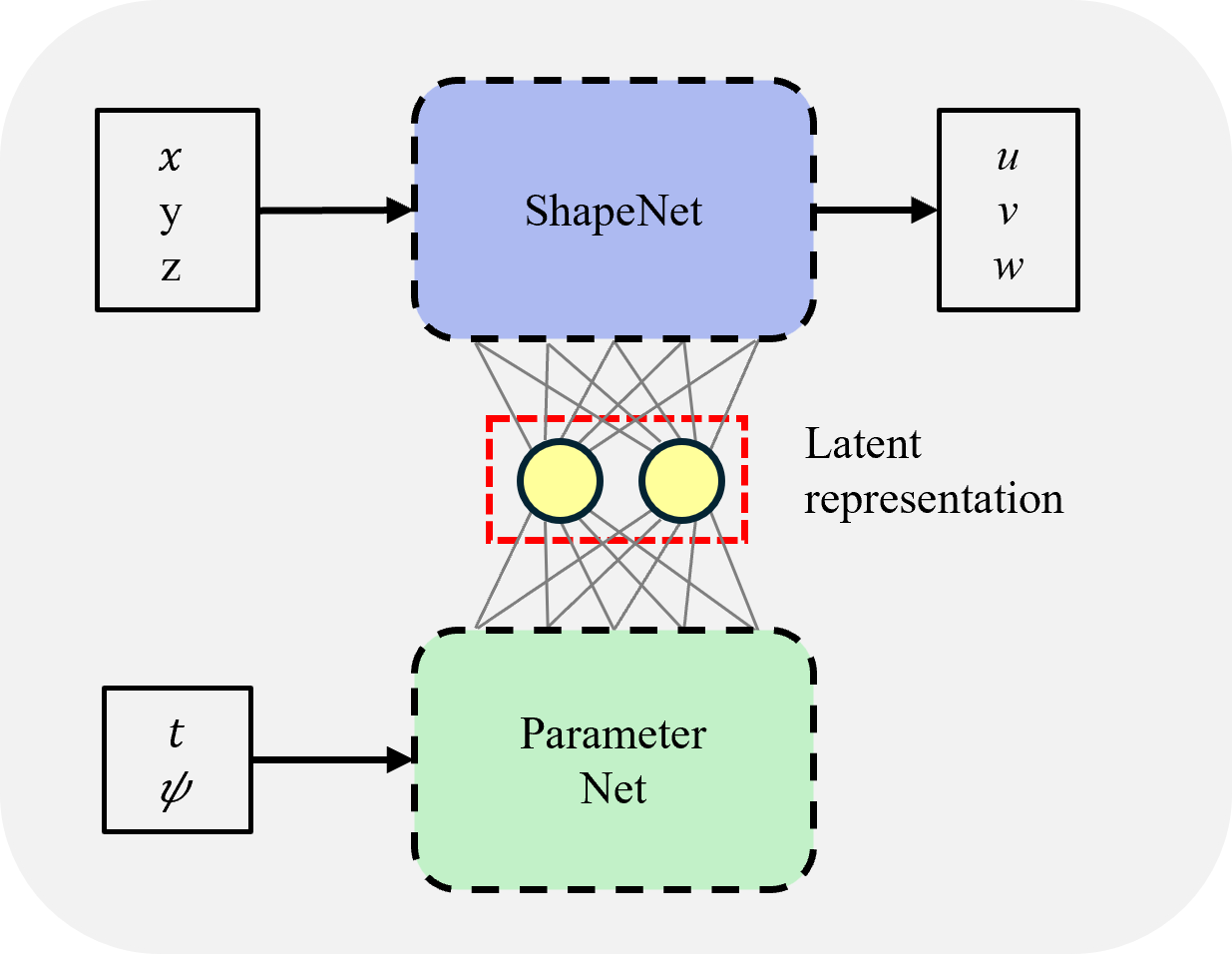}
        \caption{Original NIF architecture from Pan et al.~\cite{Pan2023}.}
        \label{fig:1}
\end{figure}

The proposed framework is designed to reconstruct fine-grained solutions of multiple scalar fields from their corresponding coarse-grained representations and constitutes an extension of the original Neural Implicit Flow (NIF) framework. Figure~\ref{fig:1} illustrates the original NIF architecture proposed by Pan et al.~\cite{Pan2023}. The network is formulated as a mesh-agnostic model for reconstructing spatio-temporal flow fields, shown here in terms of the velocity components $u$, $v$, and $w$. In the NIF architecture, the ParameterNet functions as a hypernetwork that generates a latent representation, which is subsequently used to produce the parameters of the ShapeNet dynamically. The ShapeNet is responsible for capturing the spatial structure of the target variables ($u$, $v$, and $w$) as continuous functions of the spatial coordinates $x$, $y$, and $z$, independent of any prescribed mesh. Consequently, the ShapeNet captures the spatial complexity of the solution, while the ParameterNet encodes temporal and other parametric dependencies, such as Reynolds number variations or sensor measurements. The resulting latent representation encapsulates the contribution of these dependencies within a compact latent space.

Pan et al.~\cite{Pan2023} demonstrated the versatility of the NIF architecture across four mesh-agnostic learning tasks: surrogate modeling of parametric partial differential equations, compressed representation of spatio-temporal data, learning of linear representations, and nonlinear sparse sensing.

\begin{figure}[t]
        \centering
        \includegraphics[width=0.8\textwidth]{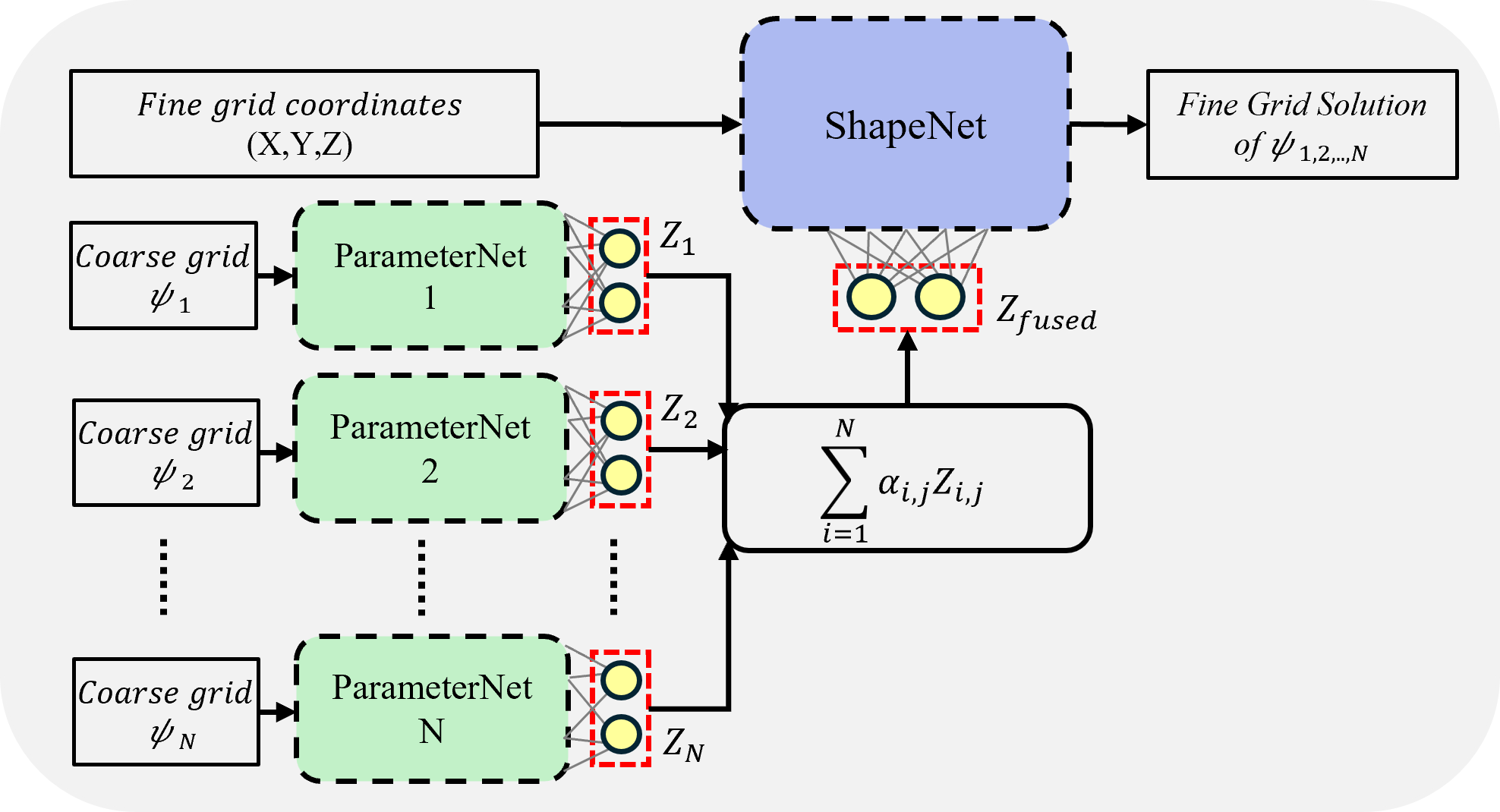}
        \caption{Proposed multi-ParameterNet NIF model for super-resolution.}
        \label{fig:2}
\end{figure}

In the proposed framework, as illustrated in Fig.~\ref{fig:2}, important changes to the NIF architecture of Pan et al.~\cite{Pan2023}. The most important change involves the inputs and outputs to the ShapeNet and the ParameterNet. The input to the ShapeNet corresponds to coordinates on which the fine-grid resolution for the scalars is constructed. The input to the ParameterNet(s) is the coarse-grid resolution of these scalars. This input replaces the time variable with a rendering of the solution field, albeit coarse, at a given time in the solution's evolution. Therefore, the coarse-grid resolution scalar fields represent the temporal evolution of these fields. A similar strategy was adopted by Amarathunga and Echekki~\cite{Amarathunga2026} with an NIF architecture to accelerate chemistry integration in combustion. In the present implementation, the scalars correspond to NLPCs that can be used to reconstruct the species and temperature in a given mechanism. The use of NLPCs is a measure to reduce the number of variables to be reconstructed {\it {via}} super-resolution, and accordingly, reduce the network size and complexity.

The second important implementation is the use of multiple ParameterNets. Each ParameterNet independently learns its own latent representation (i.e., the NLPCs) from the corresponding coarse-grained scalar inputs. These latent representations are subsequently combined into a unified latent space using a softmax-based weighting mechanism, which identifies the relative importance of each latent representation. Another key implementation is the subdomain-based learning, discussed in Sec.~\ref{sub-domain learning}, which attempts to reconstruct local features of the scalar fields and potentially extend the training to a wider range of problems.

In contrast to the nonlinear dimensionality reduction study presented in~\cite{Pan2023}, which considers a single non-reacting scalar field, namely density, reconstructed from sparse pointwise measurements using a single ParameterNet, the present framework introduces a structured latent decomposition for multi-scalar reconstruction. This design is explicit in the sense that the multicomponent thermochemical state is not treated as a single concatenated input to one conditioning network; instead, separate ParameterNets are used to encode scalar-specific latent representations before they are combined to condition the ShapeNet. While Pan et al.~\cite{Pan2023} demonstrated the effectiveness of NIF for dimensionality reduction in single-field settings, the original single-ParameterNet formulation used in that study was developed for that purpose and did not consider the additional conditioning challenges that arise in the joint reconstruction of multiple heterogeneous scalar fields. By assigning a dedicated ParameterNet to each scalar field, the proposed architecture enables each latent representation to specialize in capturing the corresponding physical or scalar contribution. The subsequent softmax-based fusion provides an adaptive mechanism to balance these contributions, effectively restructuring the conditioning of the ShapeNet rather than merely compressing the input data. This distinction is particularly important for multi-scalar reconstruction tasks, where decoupling latent roles improves parameter efficiency and enhances reconstruction fidelity.

The combined latent representation is then used to condition the ShapeNet. While the ParameterNets take coarse-grained scalar fields as inputs, the ShapeNet is evaluated at fine-grained spatial coordinates to reconstruct the high-resolution solution. Specifically, let $\{ \mathbf{z}_i \}_{i=1}^{N}$ denote the latent representations produced by the $N$ ParameterNets, where each $\mathbf{z}_i \in \mathbb{R}^{d}$ and $d$ denotes the latent dimension. A learnable weight matrix $\tilde{\boldsymbol{\alpha}} \in \mathbb{R}^{N \times d}$ is introduced and normalized using a softmax operation across the ParameterNets,
\begin{equation}
\alpha_{i,j}
=
\frac{\exp\left(\tilde{\alpha}_{i,j}\right)}
{\sum_{k=1}^{N} \exp\left(\tilde{\alpha}_{k,j}\right)},
\qquad
i = 1,\dots,N,\;\; j = 1,\dots,d .
\end{equation}
The fused latent representation $\mathbf{z}_{\mathrm{fused}} \in \mathbb{R}^{d}$ is then defined component-wise as
\begin{equation}
z_{\mathrm{fused},j}
=
\sum_{i=1}^{N} \alpha_{i,j}\, z_{i,j},
\qquad
j = 1,\dots,d .
\end{equation}

This modification reduces the number of parameters required in the proposed NIF architecture to achieve a specified training loss threshold by approximately $40\%$ compared to the original NIF framework employing a single ParameterNet. In particular, the number of dynamically generated parameters in the ShapeNet is reduced by approximately $25\%$. Since a significant portion of the computational cost in the NIF framework arises from the dynamic parameter generation (parameter of the ShapeNet), this reduction directly translates to improved computational efficiency. As a result, the overall training cost is reduced by approximately $30\%$.

In contrast, conditioning a single ParameterNet on concatenated coarse-grained scalar inputs was observed to require increased parameter capacity to achieve a comparable loss level, indicating reduced efficiency due to latent entanglement. Here, latent entanglement refers to the inability of the latent representation to distinctly encode contributions from multiple input scalars, resulting in coupled or overlapping feature representations. This coupling limits the model’s ability to learn variable-specific structures and leads to inefficient use of model capacity. By assigning dedicated ParameterNets to individual scalars, the proposed framework promotes a more structured latent decomposition, improving both parameter efficiency and reconstruction accuracy. Furthermore, even when the original architecture attains a similar training loss using a larger number of parameters, the proposed framework consistently yields superior reconstruction fidelity.

\subsection{Subdomain-Based Learning Strategy} \label{sub-domain learning}

To effectively capture the multiscale nature of reacting flows, a subdomain-based learning strategy is employed in the proposed framework. Instead of learning the global flow field in a single pass, the computational domain is decomposed into a set of overlapping local subdomains extracted from the high-resolution data. Each subdomain represents a localized region of the flow containing relevant thermochemical structures, such as flame fronts, shear layers, and reaction zones.

\begin{figure}[t]
\centering
\includegraphics[width=0.8\textwidth]{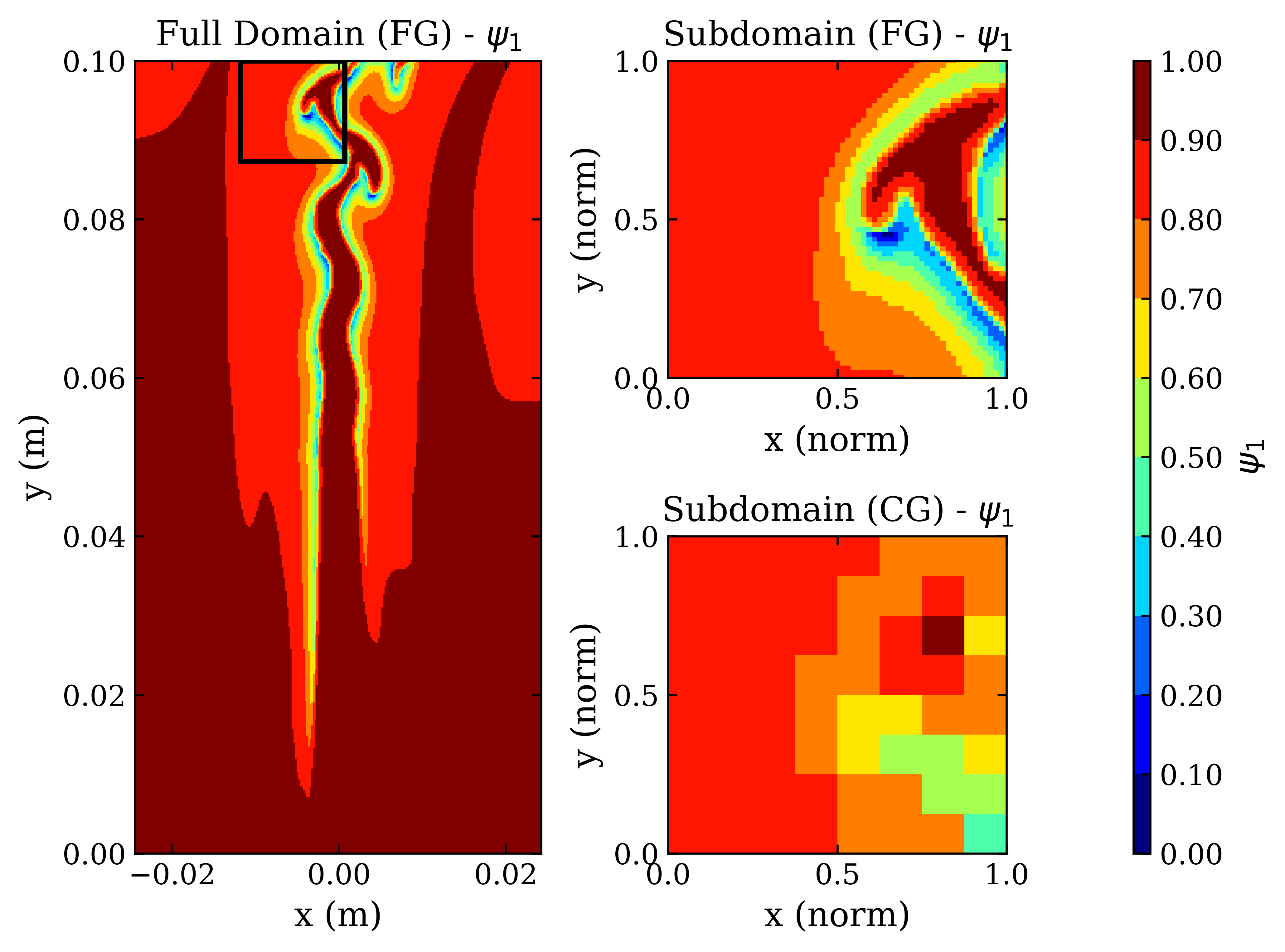}
\caption{Illustration of subdomain extraction from the normalized nonlinear principal component ($\psi_1$) from the data obtained following the procedure of section ~\ref{subsec:dataprep}. A local region is extracted from the fine-grid (FG) field and mapped to a normalized coordinate space, while the corresponding coarse-grid (CG) representation is obtained through uniform subsampling.}
\label{fig:subdomain_extraction}
\end{figure}

Figure~\ref{fig:subdomain_extraction} illustrates an example of this subdomain extraction process using the normalized nonlinear principal component ($\psi_1$), obtained following the procedure described in Section~\ref{subsec:dataprep}. The figures in the present study, displayed as pixelated images to avoid any interpolation effects by the plotting software. A localized region is selected from the full fine-grid (FG) field, and the corresponding subdomain is mapped to a normalized coordinate space. In addition, a coarse-grid (CG) representation is obtained by uniformly sampling the fine-grid solution using a prescribed coarsening factor. This results in a paired dataset consisting of a coarse input and its corresponding fine-resolution target within the same spatial region.

For each subdomain, the NIF model is trained to reconstruct the fine-resolution solution from the associated coarse-grained input. During this process, the spatial coordinates within each subdomain are normalized to a local reference frame, typically scaled to a fixed range. This normalization removes the dependence on the absolute spatial location and enables the model to learn features based on relative spatial variations rather than global positioning.

This localized formulation provides several key advantages. First, it promotes localized feature learning by allowing the model to focus on capturing fine-scale structures within a confined region, which is particularly important for reacting flows characterized by steep gradients and thin reaction zones. Second, the use of normalized local coordinates enables domain-independent learning, such that similar flow features appearing at different spatial locations or time instances are represented consistently within the learned mapping. As a result, the model does not memorize global configurations but instead learns reusable representations of fundamental flow structures.

Furthermore, the decomposition into overlapping subdomains increases the diversity of training samples by exposing the model to a wide range of localized flow patterns. This improves the robustness and generalization capability of the learned representation. Importantly, because the model learns localized and position-invariant features, the trained framework can be transferred to other configurations exhibiting similar flame structures without requiring retraining on the full domain. This makes the approach particularly suitable for complex reacting flow problems where similar physical phenomena occur across different spatial regions and operating conditions.

Finally, the global high-resolution field is reconstructed by aggregating the predictions from individual subdomains. The use of overlapping regions ensures smooth transitions between neighboring subdomains and reduces reconstruction artifacts, leading to a coherent representation of the full domain.

\subsection{Dataset Description - Direct Numerical Simulation (DNS)}\label{subsec:dns}
The proposed NIF framework for super-resolution is demonstrated on a 2D unsteady premixed Bunsen flame of methane. The DNS data was generated through the PeleLMeX open-source code \cite{Pele2024} using the DRM19 mechanism~\cite{DRM19kazakov_drm} with 19 species and 84 reversible reactions. The example source code of "PremBunsen2D" provided in the Pele distribution is used for problem formulation. The example source code was modified to include inlet jet forcing to make the flow more complex by having both temporal and spatial forcing. The inlet jet velocity used for the simulation is denoted as follows.

\begin{equation}
S(x) =
\begin{cases}
1 - \left(\dfrac{|x - x_s|}{W/2}\right)^2, & |x - x_s| \leq \dfrac{W}{2}, \\[6pt]
0, & \text{otherwise}
\end{cases}
\end{equation}

\begin{equation}
u'(x,t) = A \, S(x) \, \sin(k_x x)\,\sin(\omega t)
\end{equation}

\begin{equation}
v_y(x,t) = V_{\text{in}} \, S(x) + u'(x,t)
\end{equation}

\begin{equation}
v_y^{\text{inlet}}(x,t) = \max\left(0,\; v_y(x,t)\right)
\end{equation}

Here, $x$ and $t$ denote the spatial coordinate and time, respectively, and $x_s$ is the slot center location. 
$W$ is the slot width, and $S(x)$ defines a parabolic inlet profile. $V_{\text{in}}$ is the prescribed (user-specified) inlet velocity. 
The perturbation $u'(x,t)$ has amplitude $A$, streamwise wavenumber $k_x$, and angular frequency $\omega$. 
The quantity $v_y(x,t)$ is the unbounded inlet velocity, and $v_y^{\text{inlet}}(x,t)$ is the final inlet velocity after applying the non-negativity constraint.\par

In the present work, $V_{\text{in}}$ and $W$ are chosen such that the Reynolds number is approximately 8,000, where $W = 6~\mathrm{mm}$ denotes the full slot width. 
The values of $k_x$ and $\omega$ are selected to prescribe a controlled inlet perturbation with a characteristic wavelength and frequency. 
For $k_x = 5,000~\mathrm{m^{-1}}$, the corresponding wavelength is 
$\lambda_x = 2\pi/k_x \approx 1.26~\mathrm{mm}$, which is approximately 
$0.21W$. 
The angular frequency $\omega = 475~\mathrm{s^{-1}}$ corresponds to 
$f = \omega/(2\pi) \approx 75.6~\mathrm{Hz}$. 
Thus, the perturbation introduces a spatially and temporally periodic inlet disturbance intended to seed shear-layer instabilities while preserving the prescribed mean inlet profile.

The computational domain is two-dimensional, spanning 
$x \in [-0.025,\,0.025]~\mathrm{m}$ and $y \in [0,\,0.10]~\mathrm{m}$. 
The domain is discretized on a uniform grid with a resolution of $249 \times 505$. The characteristic flow-through time is defined as 
$t_f = L_y / V_{\text{in}} \approx 5~\mathrm{ms}$, where $L_y = 0.10~\mathrm{m}$ is the domain height. 
The simulation is advanced up to $30~\mathrm{ms}$ with a time step size of $10^{-6} s$, corresponding to approximately six flow-through times, to allow initial transients to decay. The data are then collected over the time interval $t = 30$--$30.4~\mathrm{ms}$ at intervals of $0.01~\mathrm{ms}$, yielding a total of 41 snapshots. For training purposes, every second snapshot (i.e., at intervals of $0.02~\mathrm{ms}$) is utilized.

\subsection{Training Data Preparation}\label{subsec:dataprep}

Data extracted from the time snapshots are first processed by applying a mass fraction threshold of \(10^{-4}\) to all species, excluding N$_2$ and Ar, to remove species with negligible contributions. Based on this criterion, 14 species are retained: H$_2$, H, O, O$_2$, OH, H$_2$O, HO$_2$, CH$_3$, CH$_4$, CO, CO$_2$, CH$_2$O, C$_2$H$_4$, and C$_2$H$_6$. A principal component analysis (PCA) was performed on the scalar set, including temperature and the selected species mass fractions, to identify a reduced set of representative scalars. The relative contribution of each scalar to the variance captured by the principal components was evaluated, and the scalars were ranked accordingly. Applying a cumulative contribution threshold of $85\%$ resulted in the selection of eight scalars: temperature, H$_2$, H, O$_2$, H$_2$O, CH$_4$, CO, and CO$_2$. In addition, the H radical is explicitly retained to represent thin reaction zones, where its concentration plays a critical role in capturing localized chemical activity. Notably, Kumar and Echekki~\cite{Kumar2023} employed an identical set of representative scalars in their study on accelerating turbulent combustion DNS of a methane Bunsen flame, thereby providing validation for the present selection. A similar strategy was adopted by Kumar and Echekki~\cite{Kumar2024} and Amarathunga and Echekki~\cite{Amarathunga2026} for chemistry acceleration.

\begin{figure}[t]
\centering
\includegraphics[width=0.8\textwidth]{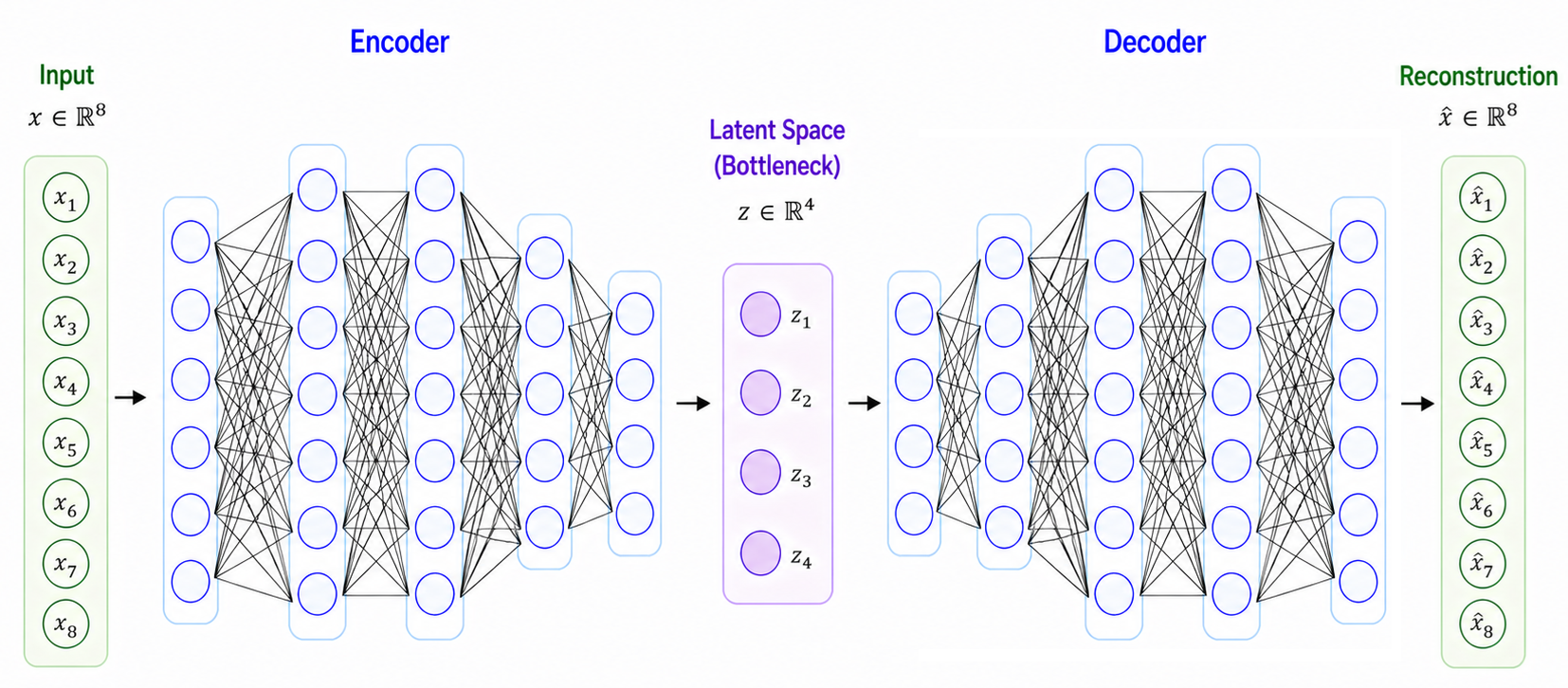}

\caption{Illustration of the Autoencoder framework for compression of representative scalars. The number of neurons and hidden layers is discussed in the text.}
\label{fig:autoencoder}
\end{figure}

These representative scalars are further subjected to dimensionality reduction using an autoencoder framework. In this approach, the eight representative scalars are compressed into a lower-dimensional latent space consisting of four nonlinear principal components (NLPCs), obtained by specifying a latent dimension of four neurons. This latent dimension was selected empirically to balance compression efficiency and reconstruction fidelity. In particular, four latent variables were found to provide sufficient reduction of the scalar dimensionality without introducing a significant loss in reconstruction accuracy. As illustrated in Figure~\ref{fig:autoencoder}, the encoder maps the input scalars to this reduced representation, while the decoder reconstructs the original scalars from the latent variables.

This procedure is conceptually analogous to principal component analysis (PCA) when no activation functions are used~\cite{PCA-linear}, in which case the autoencoder reduces to a linear transformation of the input variables that approximates the PCA subspace. However, unlike PCA, which is inherently linear, the autoencoder can incorporate nonlinear activation functions, thereby enabling nonlinear mappings between the original and reduced spaces. As a result, the latent variables obtained from the autoencoder are NLPCs, which will be denoted by $\psi$ symbol. For a fixed reduced dimensionality (i.e., the same number of latent variables as retained principal components), the autoencoder consistently provides more accurate reconstruction of the original scalars due to its ability to capture nonlinear correlations among variables~\cite{PCA-linear}.

The architecture of the autoencoder is designed to provide a symmetric encoder--decoder structure. The encoder network maps the input from 8 to 4 dimensions through a series of fully connected layers with sizes (8, 35, 50, 50, 35, 4), where 8 corresponds to the number of input representative scalars, and 4 denotes the latent variables. Notably, the intermediate expansion to higher-dimensional layers (35 and 50 neurons) enables the network to project the input into a richer feature space, thereby facilitating the learning of complex nonlinear relationships before compression into the latent space. Nonlinear activation is introduced using the Swish activation function across all hidden layers to enhance the network’s representational capacity. The decoder mirrors this structure, reconstructing the original scalars from the latent variables using layers of sizes (4, 35, 50, 50, 35, 8). This symmetric design, combined with the nonlinear expansion and subsequent compression, improves the network’s ability to accurately reconstruct the original variables while preserving their underlying nonlinear correlations. The number of hidden layers and neurons is chosen by trial-and-error, even though potentially other network optimization schemes could have been adopted.

Once the $\psi$'s are obtained from the encoder, a preprocessing step is applied prior to training. First, a constant shift is introduced to ensure that all $\psi$ values are non-negative. Subsequently, a power transformation is performed by raising the shifted values to the one-fifth power, which compresses the dynamic range and reduces the spread of the data. This transformation mitigates the dominance of large-magnitude values and leads to a more balanced representation of the $\psi$'s, thereby facilitating more effective learning by the NIF model. The transformed $\psi$'s are subsequently normalized via min–max scaling before being used for model training.

As described in Section~\ref{sub-domain learning}, once the normalized nonlinear principal components $\psi$'s) are obtained, spatial subdomains are extracted from each time snapshot. A subdomain size of $65 \times 65$ grid points (corresponding to $64 \times 64$ cells) is used, with an overlap ratio of $50\%$ between adjacent subdomains. This procedure yields 105 subdomains per snapshot and a total of 2,205 subdomains across all training snapshots. Each subdomain represents a localized spatial patch of the normalized $\psi$ fields derived from the reduced-order representation and contains information corresponding to the four retained $\psi$ variables.

To construct the corresponding coarse-grid input, the fine-grid solution is uniformly subsampled. Specifically, $32 \times 64$ grid points (corresponding to $31 \times 63$ coarse cells) are selected from the global $249 \times 505$ grid. Within each subdomain, this results in a $9 \times 9$ grid of points (corresponding to $8 \times 8$ coarse cells), representing an effective coarsening factor of 8 in both spatial directions.

While the use of overlapping subdomains increases the diversity of local scalar structures represented in the dataset, it also introduces redundancy, as similar spatial patterns may appear both within a single snapshot and across different time snapshots. To mitigate this redundancy, a feature extraction and clustering procedure is employed. Each subdomain is passed through a convolutional neural network (CNN)-based feature extractor to obtain a compact representation. The feature extractor consists of two convolutional layers with $3 \times 3$ kernels and ReLU activations, followed by spatial average pooling and a fully connected layer, resulting in a feature vector of length 64 for each subdomain. It is important to emphasize that this CNN is not trained as part of the present framework; instead, it is used as a fixed transformation to extract descriptive features from the subdomains.

The extracted feature vectors are subsequently clustered using the $k$-means algorithm to group similar subdomains. A total of 1,100 clusters are formed, and one representative subdomain is selected from each cluster to construct the final training dataset. This procedure reduces the dataset size by approximately $50\%$, decreasing the number of subdomains from 2,205 to 1,100 while retaining the diversity of the underlying features.

Finally, the input--output structure of the training dataset is constructed as follows. For each subdomain, the coarse-grid representation of the four $\psi$'s is first flattened and concatenated into a one-dimensional vector. This flattened coarse representation is then repeated for every fine-grid point within the corresponding subdomain to form the input features. In addition to the repeated coarse-grid information, the spatial coordinates of each fine-grid point, $(x_{\mathrm{fg}}, y_{\mathrm{fg}})$, are appended to the input vector.

Thus, for each fine-grid location within a subdomain, the input to the model is given by
\[
\left[\psi_{1\_\mathrm{CG}}, \psi_{2\_\mathrm{CG}}, \psi_{3\_\mathrm{CG}}, \psi_{4\_\mathrm{CG}}, x_{\mathrm{fg}}, y_{\mathrm{fg}} \right],
\]
where $\psi_{i\_\mathrm{CG}}$ denotes the flattened coarse-grid representation of the $i$-th nonlinear principal component.

The corresponding output is defined as the fine-grid values of the NLPCs at the queried spatial location, given by
\[
\left[\psi_1(x_{\mathrm{fg}}, y_{\mathrm{fg}}), \psi_2(x_{\mathrm{fg}}, y_{\mathrm{fg}}), \psi_3(x_{\mathrm{fg}}, y_{\mathrm{fg}}), \psi_4(x_{\mathrm{fg}}, y_{\mathrm{fg}})\right].
\]

This formulation enables the model to learn a continuous mapping from the coarse-grained representation and spatial coordinates to the fine-resolution $\psi$ fields, consistent with the neural implicit flow framework.

\subsection{Model Training Setup}\label{subsec:training}

To evaluate the effectiveness of the proposed framework, two NIF-based models are trained: (i) the original NIF model - hereafter referred to as the single-ParameterNet NIF architecture, and (ii) the proposed multi-ParameterNet NIF architecture. Both models are trained using the same dataset and identical training hyperparameters to ensure a fair comparison.

The architectural details of the two models are summarized in Table~\ref{tab:nif_training}. The number of hidden layers and neurons is determined empirically based on preliminary trial-and-error experiments. The ShapeNet architecture is kept identical for both models, consisting of 5 hidden layers with 90 neurons per layer, mapping spatial coordinates to the target scalar outputs. The key distinction lies in the ParameterNet design. The original NIF employs a single ParameterNet that takes the concatenated coarse-grid representation as input, whereas the proposed framework utilizes four independent ParameterNets, each associated with one nonlinear principal component ($\psi$).

Each ParameterNet in the proposed model has a reduced input dimensionality and learns a dedicated latent representation, which is subsequently fused as described in Section~\ref{subsec:nif}. Despite this structural modification, both models are designed to have approximately the same total number of trainable parameters to ensure a balanced comparison. 

\begin{table}
\centering
\caption{Comparison of training configurations for single-ParameterNet and multi-ParameterNet NIF models.}
\label{tab:nif_training}
\begin{tabular}{lcc}
\hline
\textbf{Configuration} & \textbf{Single-ParameterNet NIF} & \textbf{Multi-ParameterNet NIF} \\
\hline
ShapeNet -- Hidden layers \& neurons & (5, 90) & (5, 90) \\
ShapeNet -- Activation function & $\tanh$ & $\tanh$ \\
ShapeNet -- Input \& output & (2, 4) & (2, 4) \\
Number of ParameterNets & 1 & 4 \\
ParameterNet -- Hidden layers \& neurons & (5, 90) & (5, 50) per network \\
ParameterNet -- Activation function & Swish & Swish \\
ParameterNet -- Input dimension & 324 & 81 per network \\
Latent space dimension & 16 & 16 \\
Total number of parameters & $\sim$778K & $\sim$778K \\
\hline
\end{tabular}
\end{table}

Both models are trained using identical optimization settings. The training is performed for 15{,}000 epochs using an exponentially decaying learning rate schedule. The initial learning rate is set to $5 \times 10^{-3}$, with a decay step of 50 epochs and a decay rate of 0.9. A minimum (floor) learning rate of $10^{-6}$ is enforced to prevent the learning rate from vanishing during training. This consistent setup ensures that any observed differences in performance can be attributed solely to the architectural modifications rather than differences in training procedure. The hyperparameters were selected empirically through preliminary experiments to ensure stable and effective training.

%\subsection{Evaluation Metrics}

The performance of the proposed super-resolution framework is evaluated using three complementary metrics: the Mean Absolute Error (MAE), the Maximum Error (ME), and the Structural Similarity Index Measure (SSIM). All metrics are computed using the normalized thermochemical fields. While MAE and ME quantify pointwise reconstruction accuracy, SSIM evaluates the preservation of spatial structures and local patterns within the reconstructed fields.

The Mean Absolute Error is defined as

\begin{equation}
\mathrm{MAE}
=
\frac{1}{N}
\sum_{i=1}^{N}
\left| y_i^{*} - \hat{y}_i^{*} \right|,
\end{equation}

where $y_i^{*}$ and $\hat{y}_i^{*}$ denote the normalized reference and reconstructed values, respectively, and $N$ is the total number of grid points.

The Maximum Realtive Error is computed as

\begin{equation}
\mathrm{MRE}
=
\max_{i}
\left|
y_i^{*} - \hat{y}_i^{*}
\right|,
\end{equation}

Since the thermochemical fields are normalized to the range $[0,1]$, the normalization range is unity, i.e., $y_{\max}^{}-y_{\min}^{}=1-0=1$. Therefore, the maximum absolute deviation in the normalized space is equivalent to the maximum relative error. Accordingly, MRE represents the largest relative pointwise deviation between the normalized reconstructed and reference fields.

To assess the structural fidelity of the reconstructed fields, the Structural Similarity Index Measure (SSIM) is employed \cite{SSIM}. Unlike pointwise error metrics, SSIM evaluates the similarity between the reconstructed and reference fields by comparing their local mean, variance, and covariance structures. Originally introduced as an image quality assessment metric based on structural similarity rather than error visibility alone, SSIM provides a measure of how well spatial patterns and coherent structures are preserved. This property is particularly relevant for super-resolution tasks, where reconstruction quality depends not only on minimizing pointwise errors but also on retaining structural fidelity in the recovered high-resolution fields \cite{zhou2021}. In the context of turbulent-flow super-resolution, SSIM has also been adopted as an evaluation metric because its dependence on mean, variance, and covariance makes it suitable for assessing the statistical and structural characteristics of turbulent fields \cite{Blastnet3d}. Therefore, SSIM is well-suited for reacting-flow super-resolution, where accurately reconstructing flame fronts, reaction zones, species gradients, and coherent thermochemical structures is often as important as reducing local pointwise errors. Accordingly, SSIM provides a complementary assessment to MAE and MRE by quantifying the preservation of thermochemical patterns and field morphology.

The SSIM between two fields $x$ and $y$ is defined as

\begin{equation}
\mathrm{SSIM}(x,y)
=
\frac{
(2\mu_x\mu_y+C_1)(2\sigma_{xy}+C_2)
}{
(\mu_x^2+\mu_y^2+C_1)
(\sigma_x^2+\sigma_y^2+C_2)
},
\end{equation}

where $\mu_x$ and $\mu_y$ denote the mean values of the reference and reconstructed fields, respectively, $\sigma_x^2$ and $\sigma_y^2$ denote their variances, and $\sigma_{xy}$ is their covariance. The constants $C_1$ and $C_2$ are stabilization constants used to avoid numerical instabilities when the local mean or variance is small. In the present work, all thermochemical scalar fields are normalized to the range $[0,1]$, and SSIM is computed using a dynamic range of $L=1$. Following the standard SSIM formulation, the constants are defined as $C_1=(K_1L)^2$ and $C_2=(K_2L)^2$, with $K_1=0.01$ and $K_2=0.03$. Therefore, the values used here are $C_1=1.0\times10^{-4}$ and $C_2=9.0\times10^{-4}$. SSIM values range from 0 to 1, with a value of 1 corresponding to perfect structural agreement between the reconstructed and reference fields.

Together, these metrics provide a comprehensive assessment of reconstruction quality by quantifying both pointwise numerical accuracy and preservation of spatial structures in the thermochemical fields.

\section{Results and Discussion}\label{sec:results}
In this section, the performance of the multi-parameterNet NIF-based super-resolution framework is evaluated through \textit{a priori} analysis using direct numerical simulation (DNS) data of a turbulent premixed Bunsen flame. The objective is to assess the framework’s capability to reconstruct fine-resolution fields from coarse-resolution inputs while preserving key flow and combustion features. The evaluation is carried out hierarchically: first, the reconstruction accuracy of the nonlinear principal components ($\psi$'s) is examined at the subdomain level, followed by assessment over the full computational domain. Subsequently, the reconstructed $\psi$'s are mapped back to the physical space to recover temperature and species fields, enabling direct comparison with DNS data. Finally, the proposed framework is compared against the original NIF formulation to quantify improvements in reconstruction accuracy and generalization.

\subsection{Results on Subdomain}\label{subsec:sub-dmoain}

\begin{figure}[tbp]
    \centering
    \begin{subfigure}{\linewidth}
        \centering
        \includegraphics[width=0.8\linewidth]{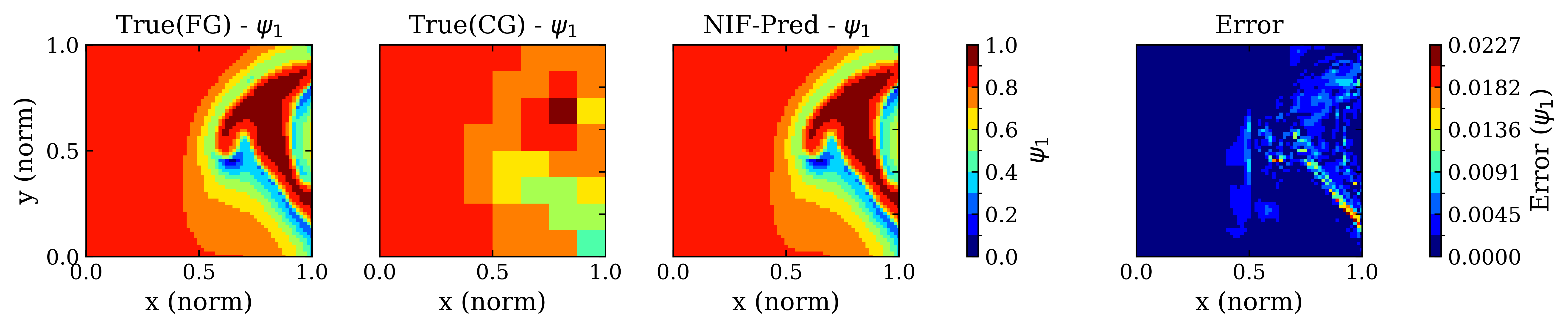}
    \end{subfigure}

    \begin{subfigure}{\linewidth}
        \centering
        \includegraphics[width=0.8\linewidth]{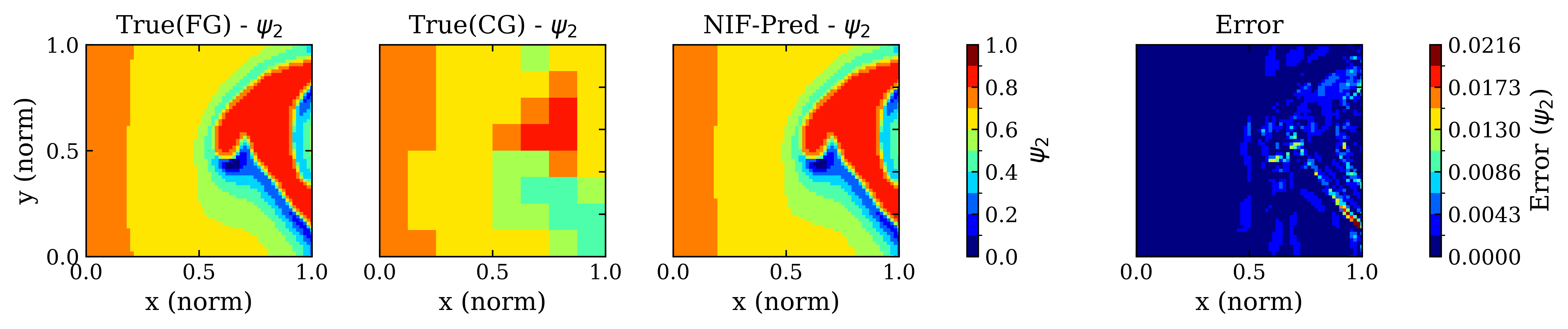}
    \end{subfigure}

    \begin{subfigure}{\linewidth}
        \centering
        \includegraphics[width=0.8\linewidth]{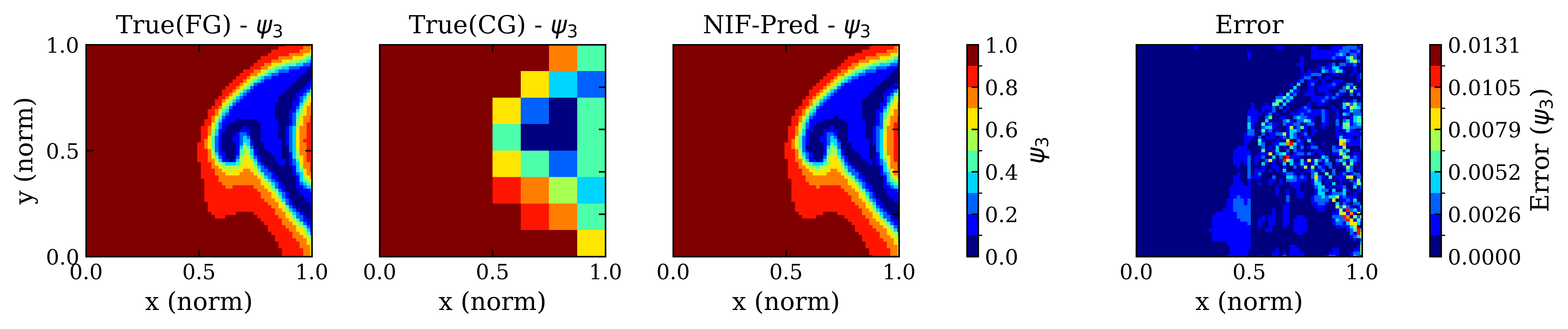}
    \end{subfigure}

    \begin{subfigure}{\linewidth}
        \centering
        \includegraphics[width=0.8\linewidth]{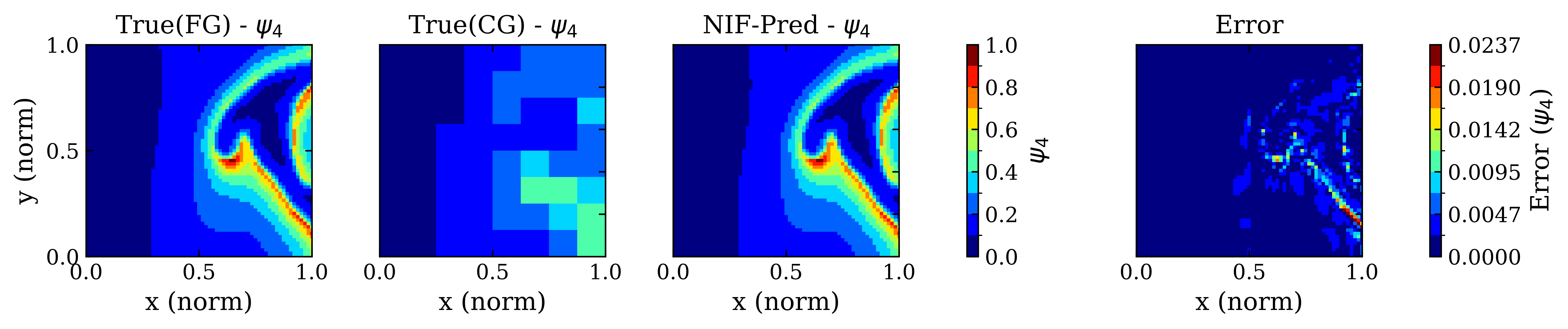}
    \end{subfigure}

    \caption{Comparison of true normalized fine grid (FG) $\psi$'s, coarse grid (CG) $\psi$'s from uniform sampling, and NIF super-resolution predictions (FG)  with absolute error between ground truth and NIF predictions for $\psi$'s on a randomly selected sub-domain.}

    \label{fig:sub-domain-nif}
\end{figure}

In this section, results are presented for the proposed multi-ParameterNet NIF framework. The reconstruction performance of the proposed NIF framework is first evaluated at the subdomain level. Figure~\ref{fig:sub-domain-nif} presents a representative example of a randomly selected subdomain, comparing the true fine-grid (FG) nonlinear principal components ($\psi$'s), the corresponding coarse-grid (CG) inputs obtained through uniform subsampling, and the NIF-predicted FG reconstructions. The absolute error between the predicted and true fields is also shown for each $\psi$.

The CG representations exhibit a significant loss of spatial detail due to the sparse nature of uniform subsampling, where only an $8 \times 8$ set of cells is retained within each subdomain. This severe reduction in resolution results in under-resolved fields that lack the fine-scale structures present in the FG data, particularly in regions of steep gradients and thin reaction zones. Despite this substantial loss of information, the NIF model accurately reconstructs the FG fields, effectively recovering sharp spatial variations and fine-scale features from the highly coarsened input.

Across all four $\psi$'s, the reconstructed fields show excellent agreement with the FG data, capturing both smooth variations and localized gradients. Although the $\psi$'s obtained from the autoencoder do not have a direct physical interpretation, the results demonstrate that the model preserves the essential spatial characteristics of the underlying fields. The reconstruction errors are primarily confined to regions of high gradients, where information loss due to subsampling is most pronounced, while the predictions in smoother regions exhibit near-perfect agreement.

Quantitatively, the reconstructions are highly accurate over the training time range, including unseen intermediate snapshots. The mean absolute errors(MAEs) are $0.08\%$, $0.06\%$, $0.07\%$, and $0.05\%$ for $\psi_{1-4}$, respectively. The maximum relative reconstruction errors(MREs), computed with respect to the range of each $\psi$, remain below $4.2\%$, with values of $3.8\%$ for $\psi_1$, $4.2\%$ for $\psi_2$, $3.0\%$ for $\psi_3$, and $3.9\%$ for $\psi_4$. These results indicate consistent spatial and temporal accuracy of the model within the training window. Overall, the proposed NIF framework provides an accurate and robust mapping from coarse-grid inputs to fine-grid $\psi$ fields at the subdomain level, forming a strong basis for full-domain reconstruction.

\subsection{Global Domain Reconstruction}\label{subsec:global domain}

\begin{figure}[tbp]
    \centering
    \begin{subfigure}{\linewidth}
        \centering
        \includegraphics[width=0.8\linewidth]{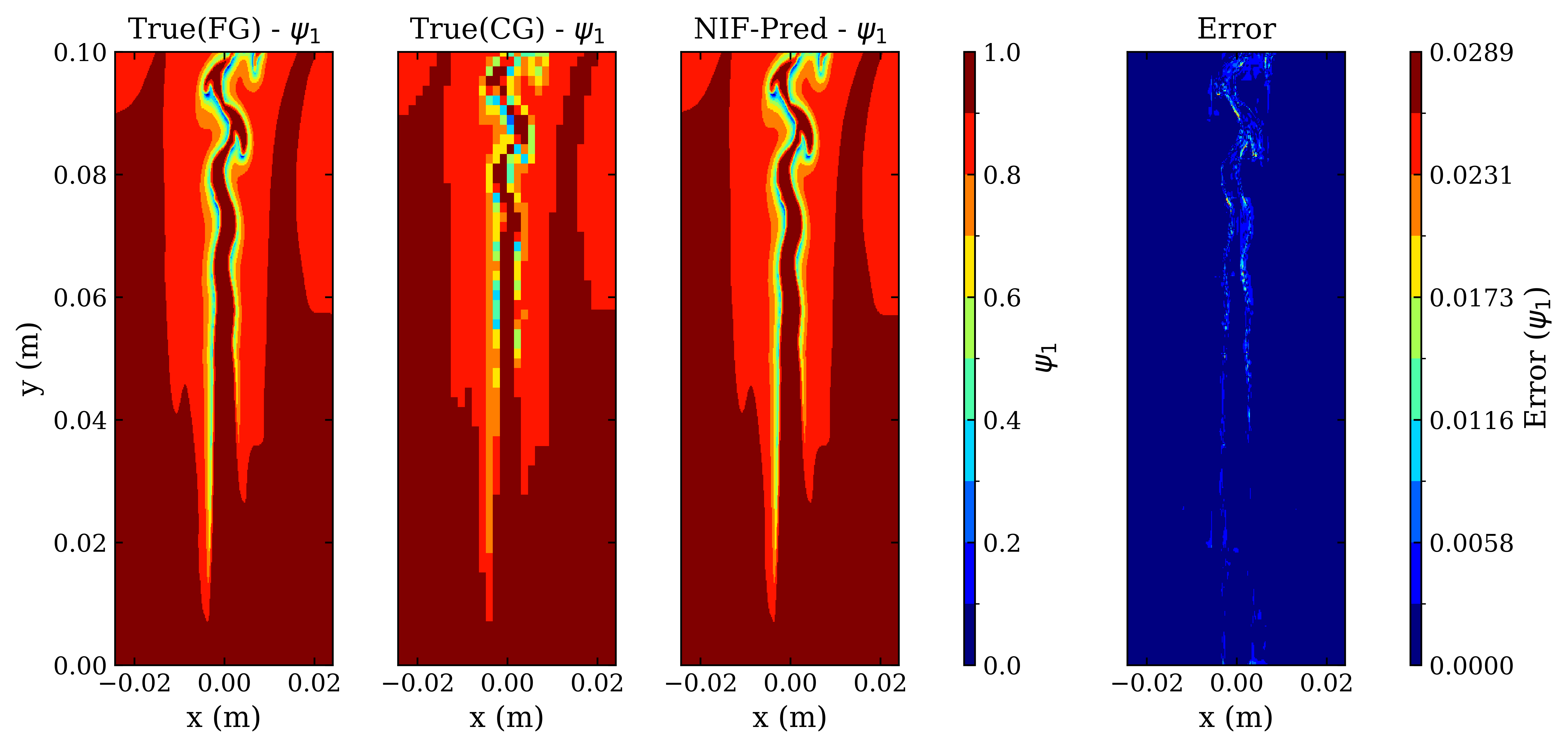}
    \end{subfigure}

    \begin{subfigure}{\linewidth}
        \centering
        \includegraphics[width=0.8\linewidth]{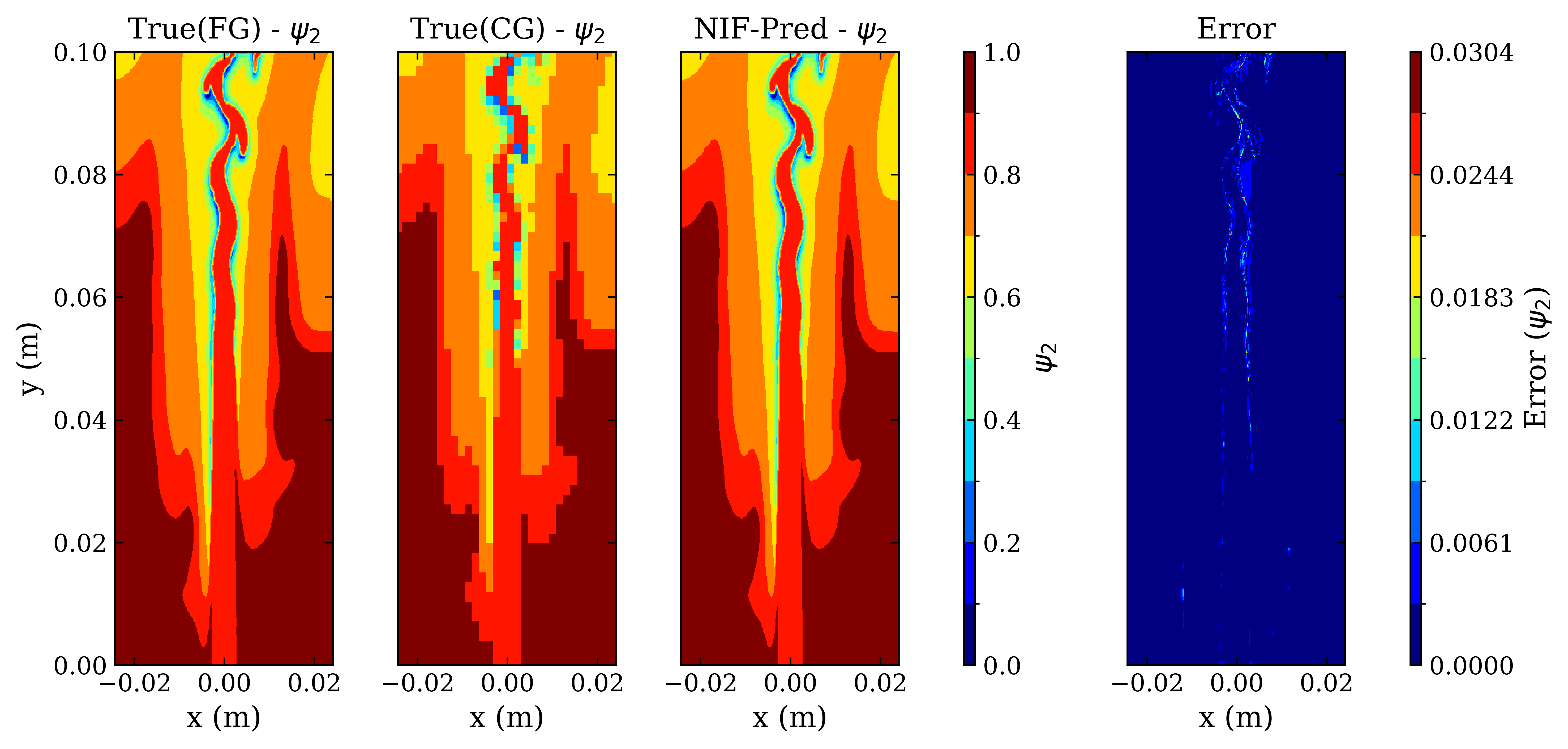}
    \end{subfigure}
    
    \caption{Comparison of normalized fine grid (FG) (ground truth $\psi$'s), coarse grid (CG), and NIF super-resolution predictions (FG)  with absolute error between ground truth and NIF predictions for $\psi_1$ \& $\psi_2$.}
    \label{fig:global-nif-pred1}
\end{figure}

\begin{figure}[tbp]
    \centering
    \begin{subfigure}{\linewidth}
        \centering
        \includegraphics[width=0.8\linewidth]{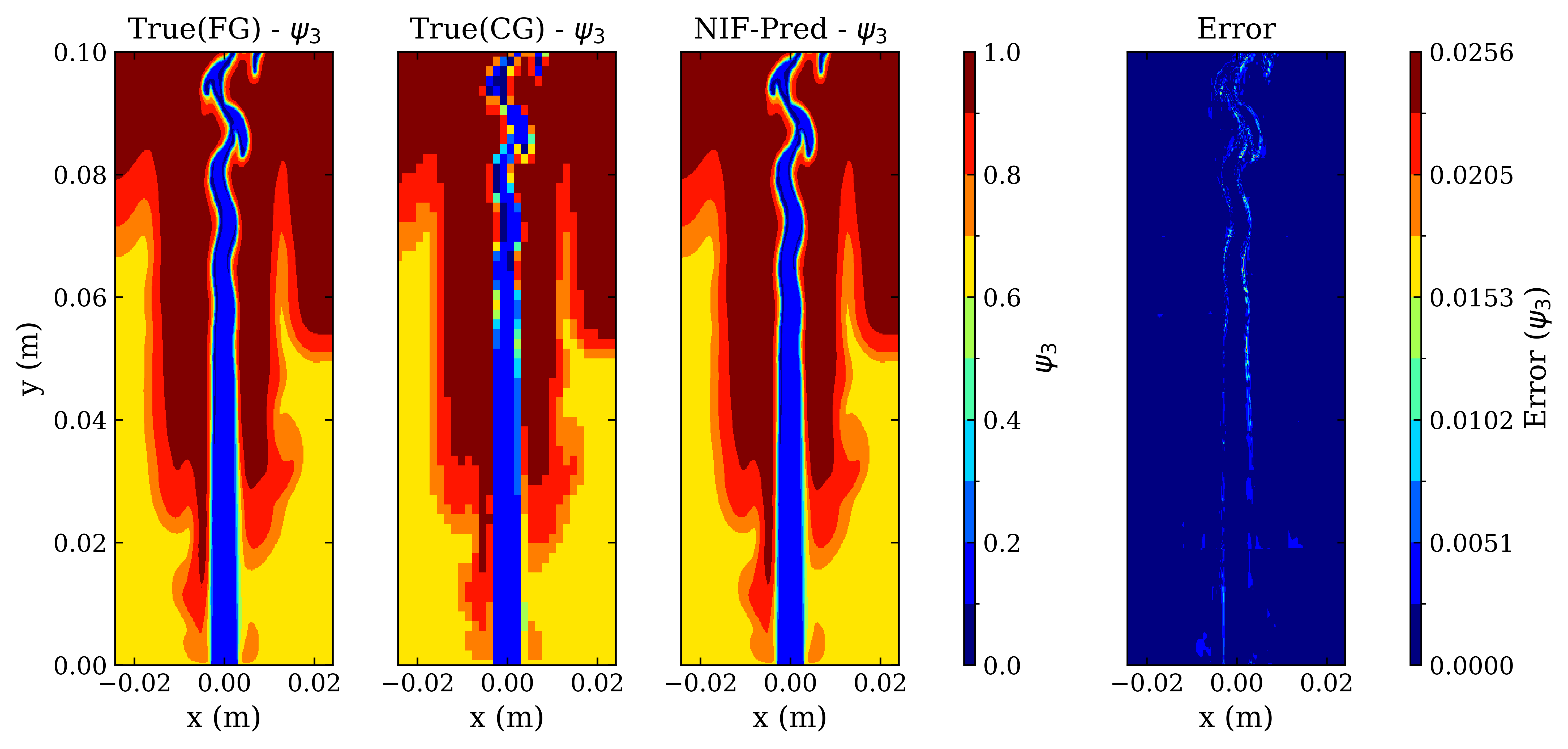}
    \end{subfigure}
    \begin{subfigure}{\linewidth}
        \centering
        \includegraphics[width=0.8\linewidth]{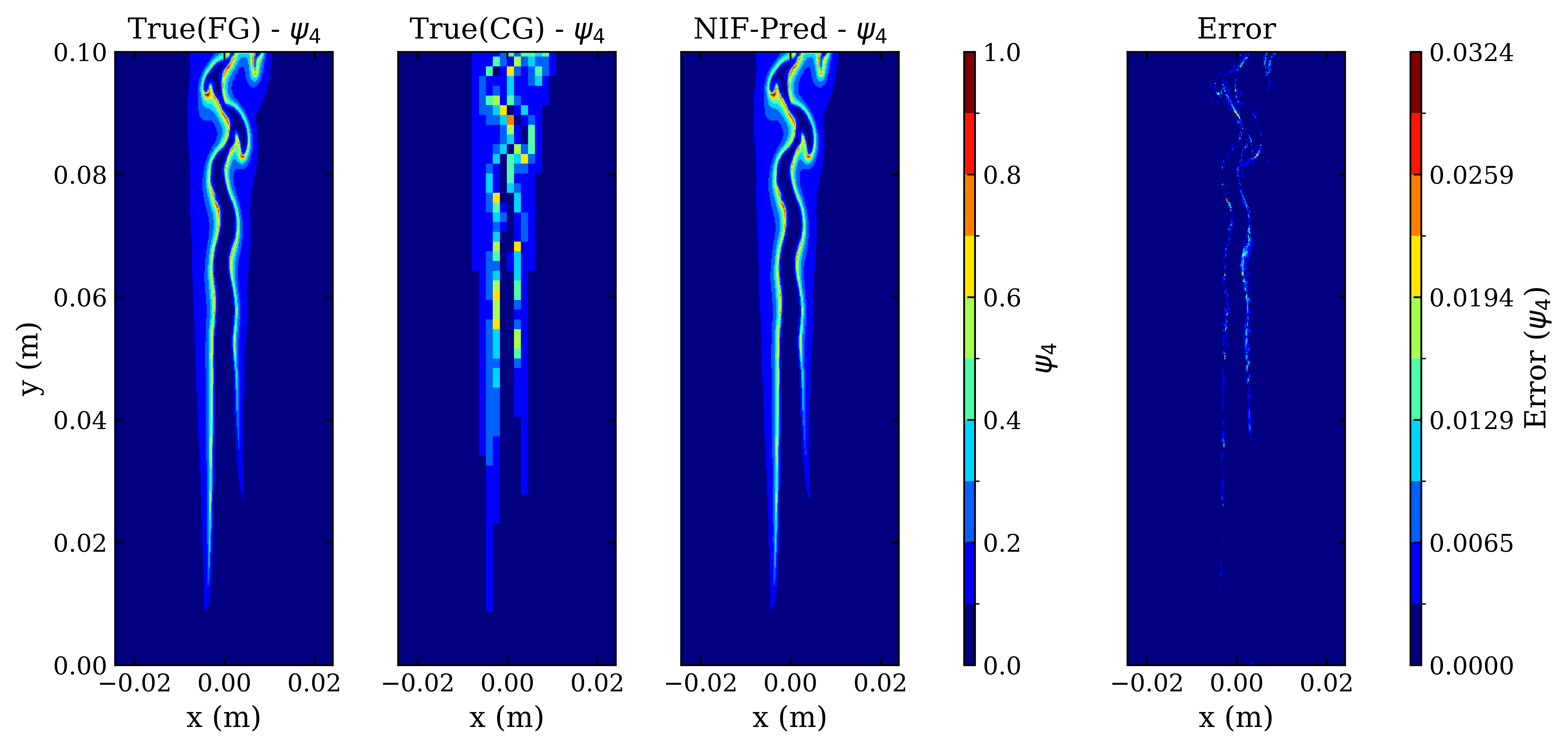}
    \end{subfigure}

    \caption{Comparison of normalized fine grid (FG) (ground truth $\psi$'s), coarse grid (CG), and NIF super-resolution predictions (FG)  with absolute error between ground truth and NIF predictions for $\psi_3$ \& $\psi_4$.}
    \label{fig:global-nif-pred2}
\end{figure}

Following the subdomain-level evaluation, the global fine-resolution field is reconstructed using predictions obtained from the proposed multi-ParameterNet NIF framework. This is achieved by aggregating outputs from overlapping subdomains. As described in Section~\ref{sub-domain learning}, the domain is decomposed into a set of overlapping subdomains with a $50\%$ overlap. Consequently, each spatial location in the global domain is covered by multiple subdomains, leading to multiple candidate predictions for the same point.

A direct averaging of overlapping predictions can introduce smoothing effects and may propagate errors from poorly reconstructed subdomains. To address this, a selection-based reconstruction strategy is employed. For each subdomain, a reconstruction error is computed on the coarse-grid locations by comparing the predicted and true values at the corresponding subsampled points. This coarse-grid MAE serves as a proxy for the overall reconstruction quality of the subdomain.

During the aggregation process, each spatial location is assigned the prediction from the subdomain that yields the lowest coarse-grid MAE among all overlapping candidates. This ensures that the most accurate local reconstruction is retained, while predictions from subdomains with higher errors are discarded. As a result, the method effectively mitigates the influence of poorly reconstructed regions, which could otherwise degrade the overall solution.

The global reconstruction is thus obtained by stitching together locally optimal predictions from overlapping subdomains, resulting in a coherent and high-fidelity representation of the fine-grid $\psi$ fields. This approach preserves sharp spatial features and avoids artificial smoothing, which is particularly important for capturing thin reaction zones and steep gradients in reacting flows.

Figures~\ref{fig:global-nif-pred1} and \ref{fig:global-nif-pred2} present comparisons between the reconstructed global fields and the corresponding fine-grid ground truth for all $\psi$'s. Each figure is organized such that, for a given $\psi$, the columns correspond to the true fine-grid (FG) field, the coarse-grid (CG) representation obtained via uniform subsampling, the NIF-reconstructed FG field, and the corresponding absolute error. The figure~\ref{fig:global-nif-pred1} displays results for $\psi_1$ and $\psi_2$, while the figure~\ref{fig:global-nif-pred2} presents $\psi_3$ and $\psi_4$. The NIF predictions closely match the true fields across the entire domain, accurately capturing both large-scale structures and localized features. The error distributions indicate that discrepancies remain confined to regions of high gradients, while the majority of the domain exhibits excellent agreement (pointwise absolute error below $1\%$).

Quantitatively, the global reconstruction preserves the accuracy achieved at the subdomain level, with no observable degradation due to the aggregation of overlapping predictions. The MAE and MRE metrics remain consistent with the subdomain results, with maximum relative errors below $4.2\%$ for all $\psi$'s (computed with respect to their respective ranges). This confirms that the selection-based stitching strategy effectively avoids error accumulation and ensures a stable and accurate reconstruction across the entire domain.

Overall, the results demonstrate that the above aggregation strategy, combined with the NIF-based subdomain reconstruction, provides an accurate and robust framework for recovering global fine-resolution fields from highly coarsened inputs.

\begin{figure}[t]
\centering
\includegraphics[width=0.7\textwidth]{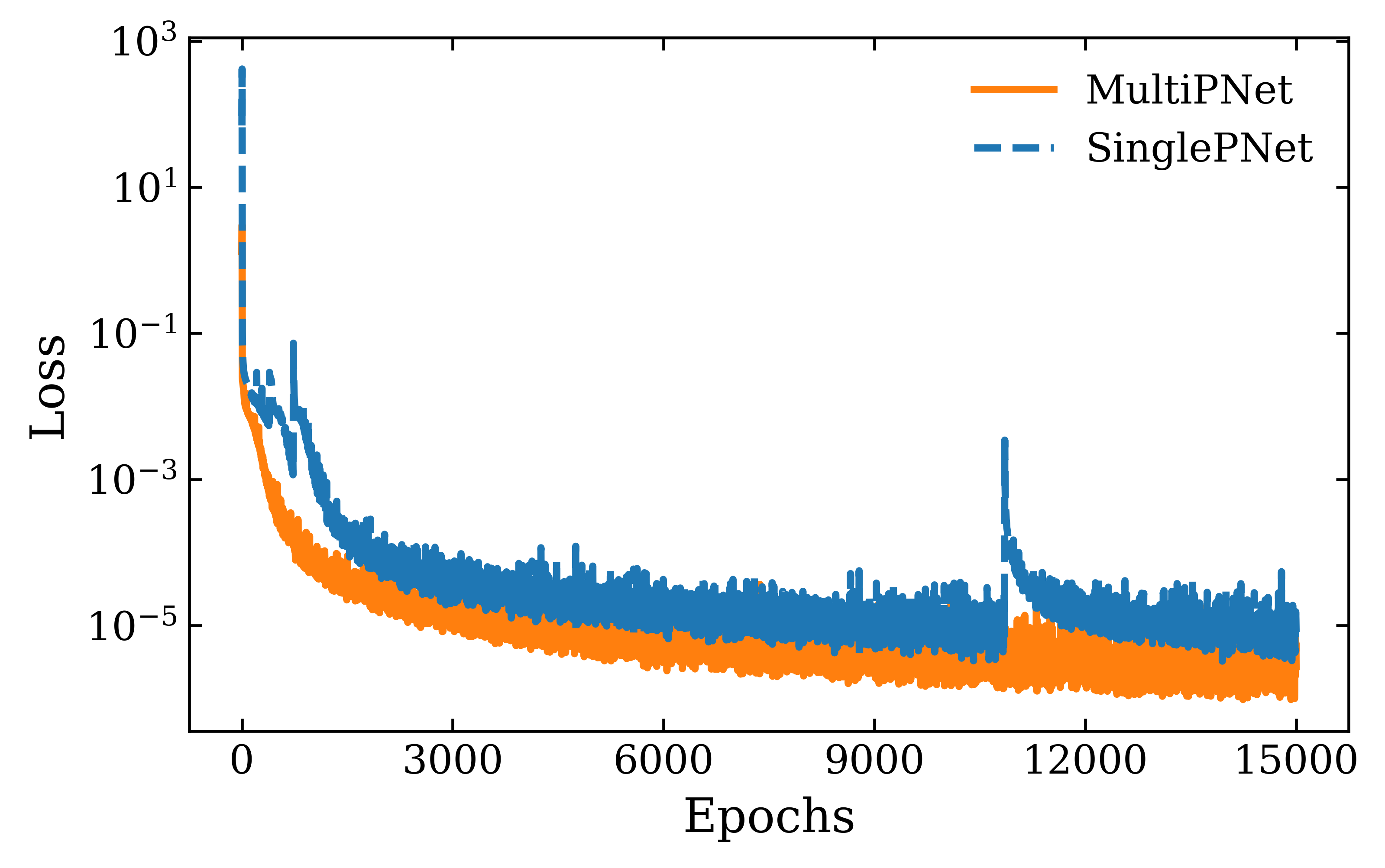}
\caption{Training loss comparison between the original NIF (SinglePNet) and the proposed NIF framework (MultiPNet).}
\label{fig:trainingloss}
\end{figure}

\begin{figure}[h!]
\centering

\begin{subfigure}{0.48\textwidth}
    \centering
    \includegraphics[width=\linewidth]{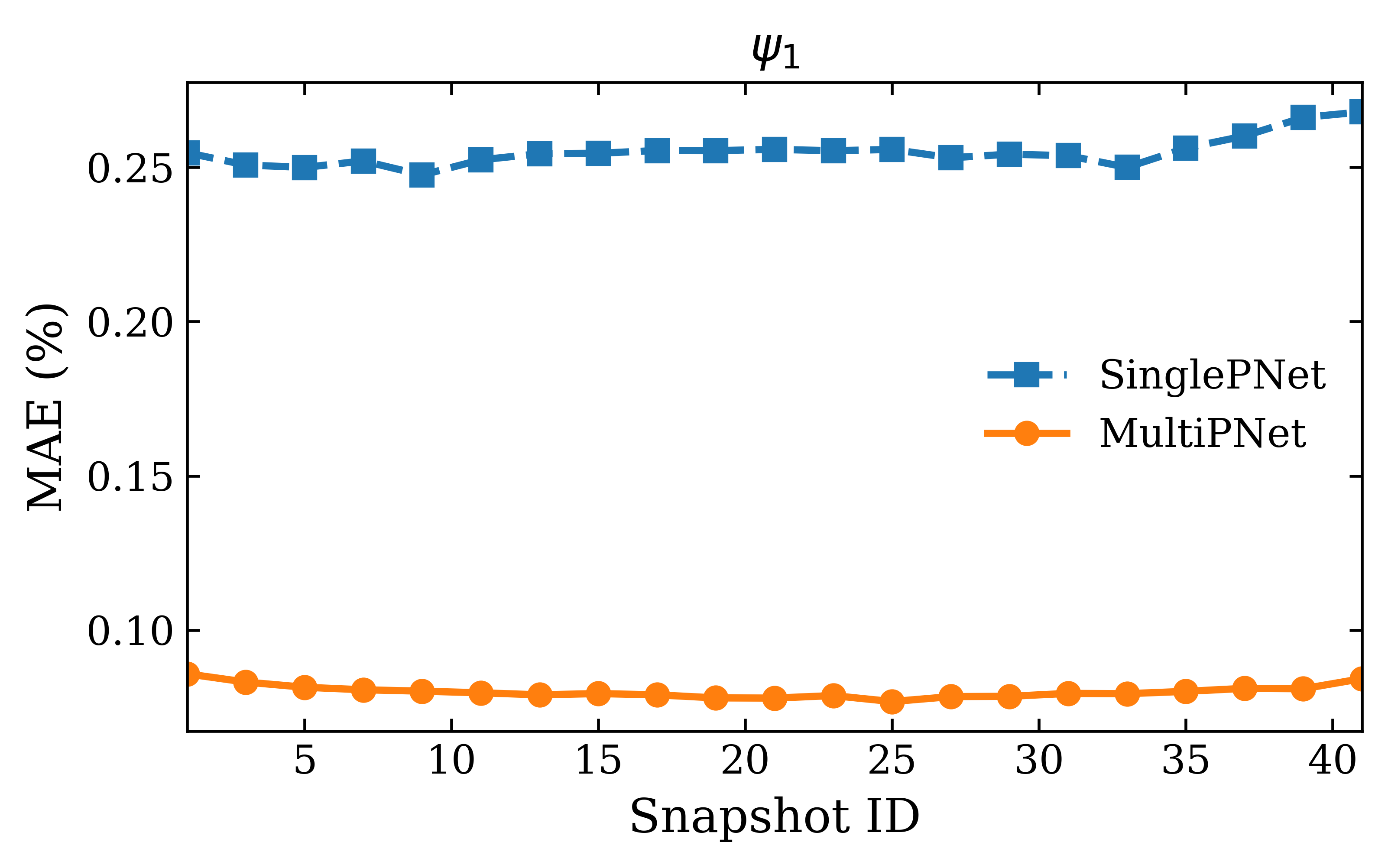}
\end{subfigure}
\hfill
\begin{subfigure}{0.48\textwidth}
    \centering
    \includegraphics[width=\linewidth]{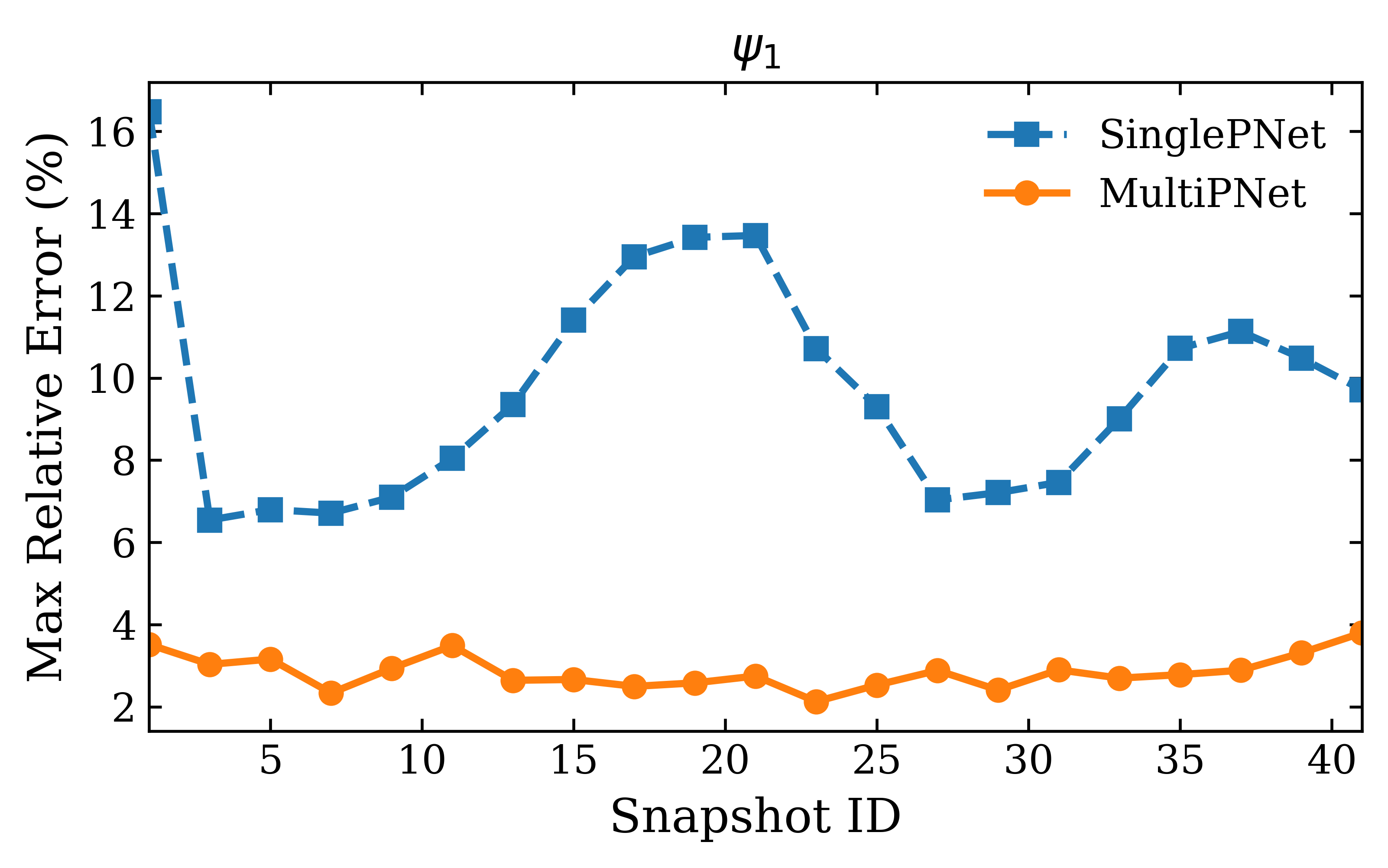}
\end{subfigure}

\vspace{0.5em}

\begin{subfigure}{0.48\textwidth}
    \centering
    \includegraphics[width=\linewidth]{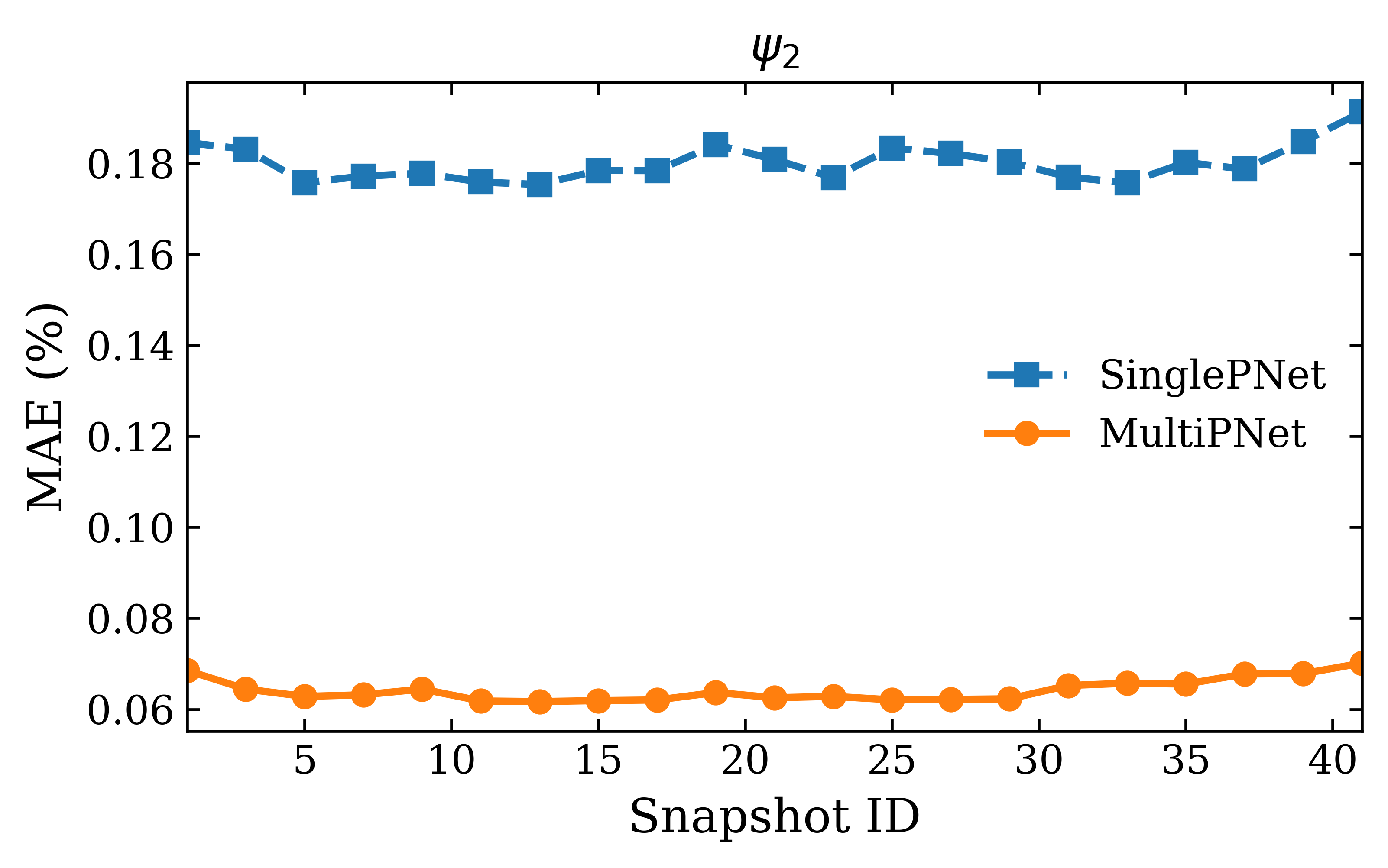}
\end{subfigure}
\hfill
\begin{subfigure}{0.48\textwidth}
    \centering
    \includegraphics[width=\linewidth]{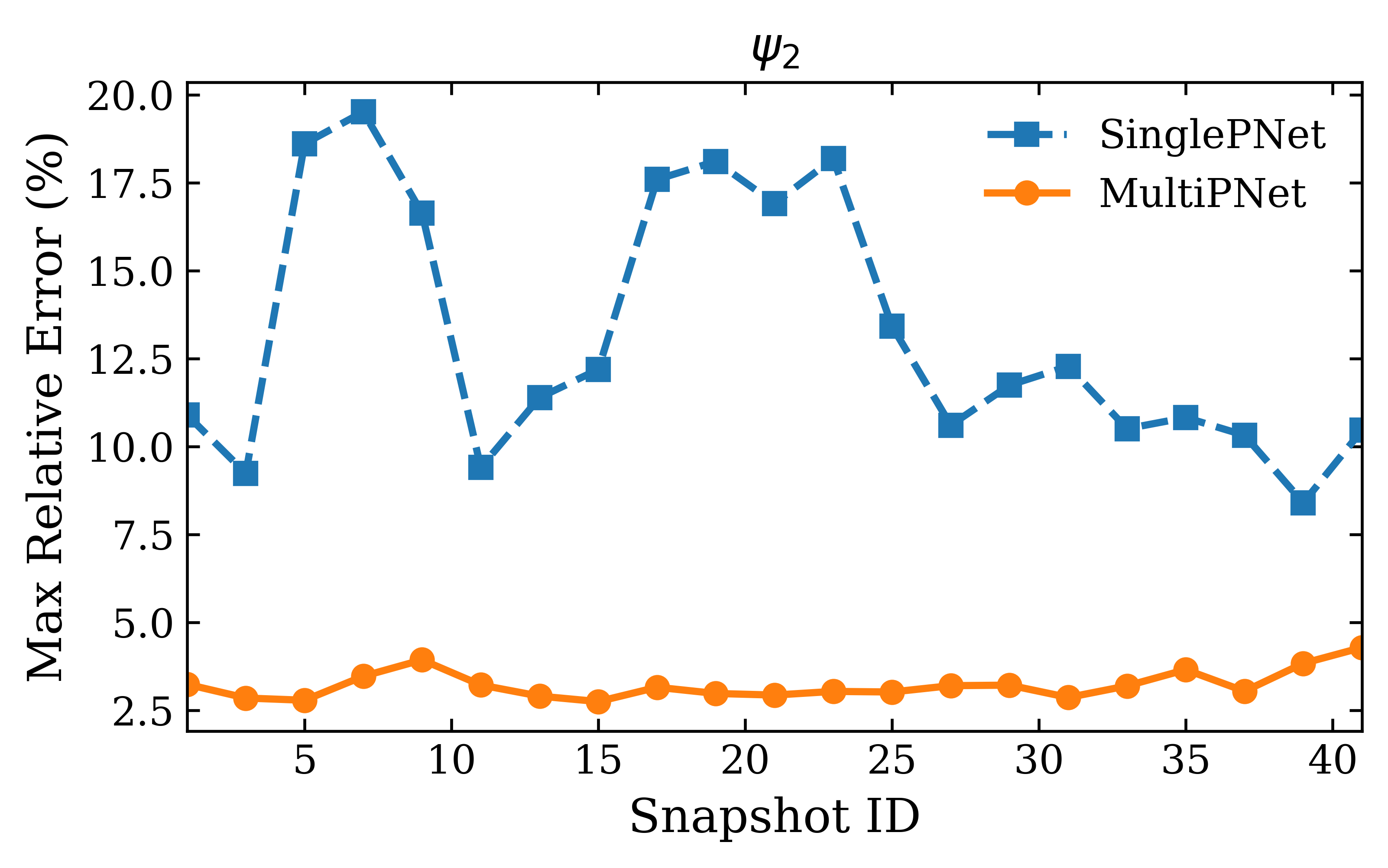}
\end{subfigure}

\vspace{0.5em}

\begin{subfigure}{0.48\textwidth}
    \centering
    \includegraphics[width=\linewidth]{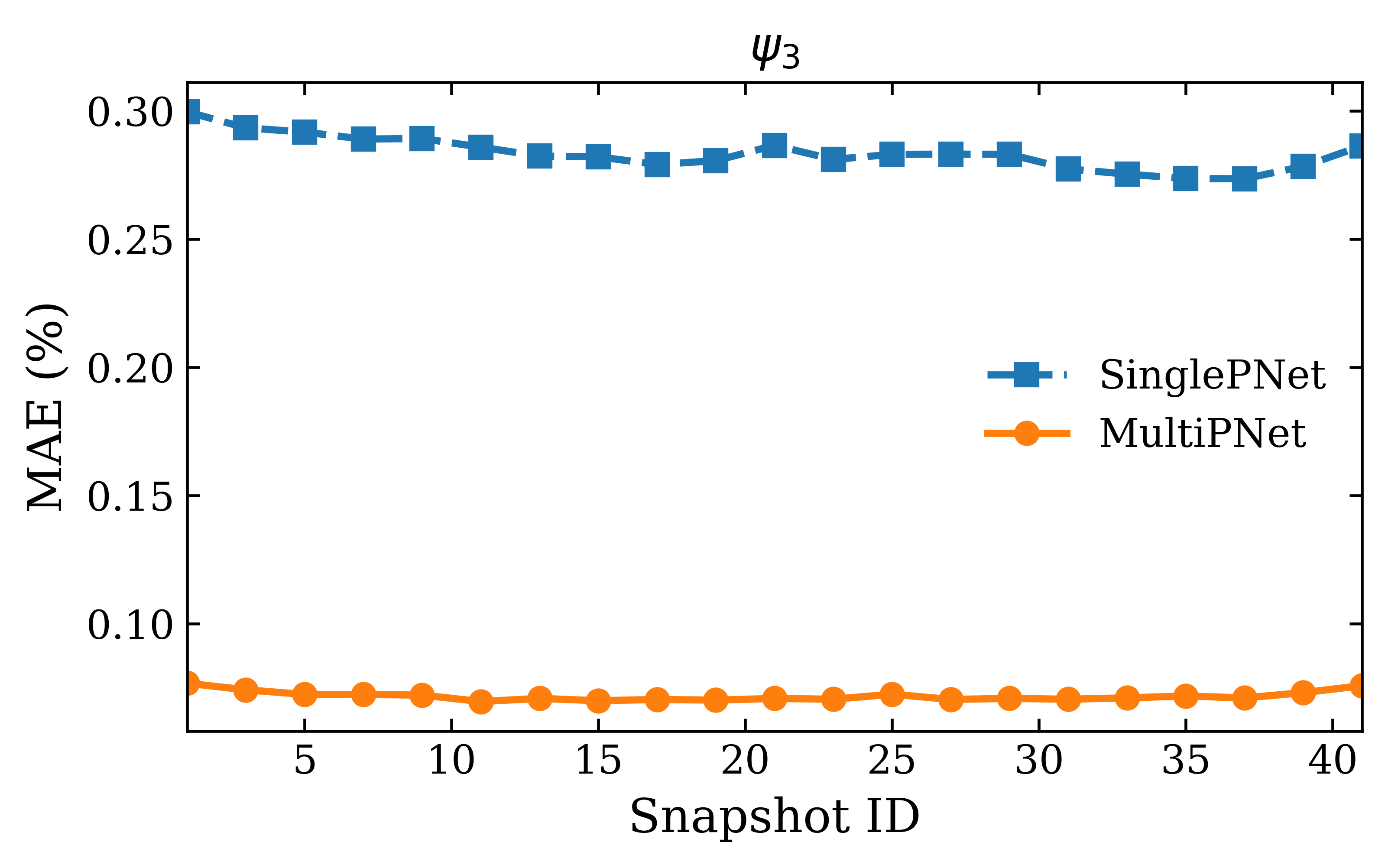}
\end{subfigure}
\hfill
\begin{subfigure}{0.48\textwidth}
    \centering
    \includegraphics[width=\linewidth]{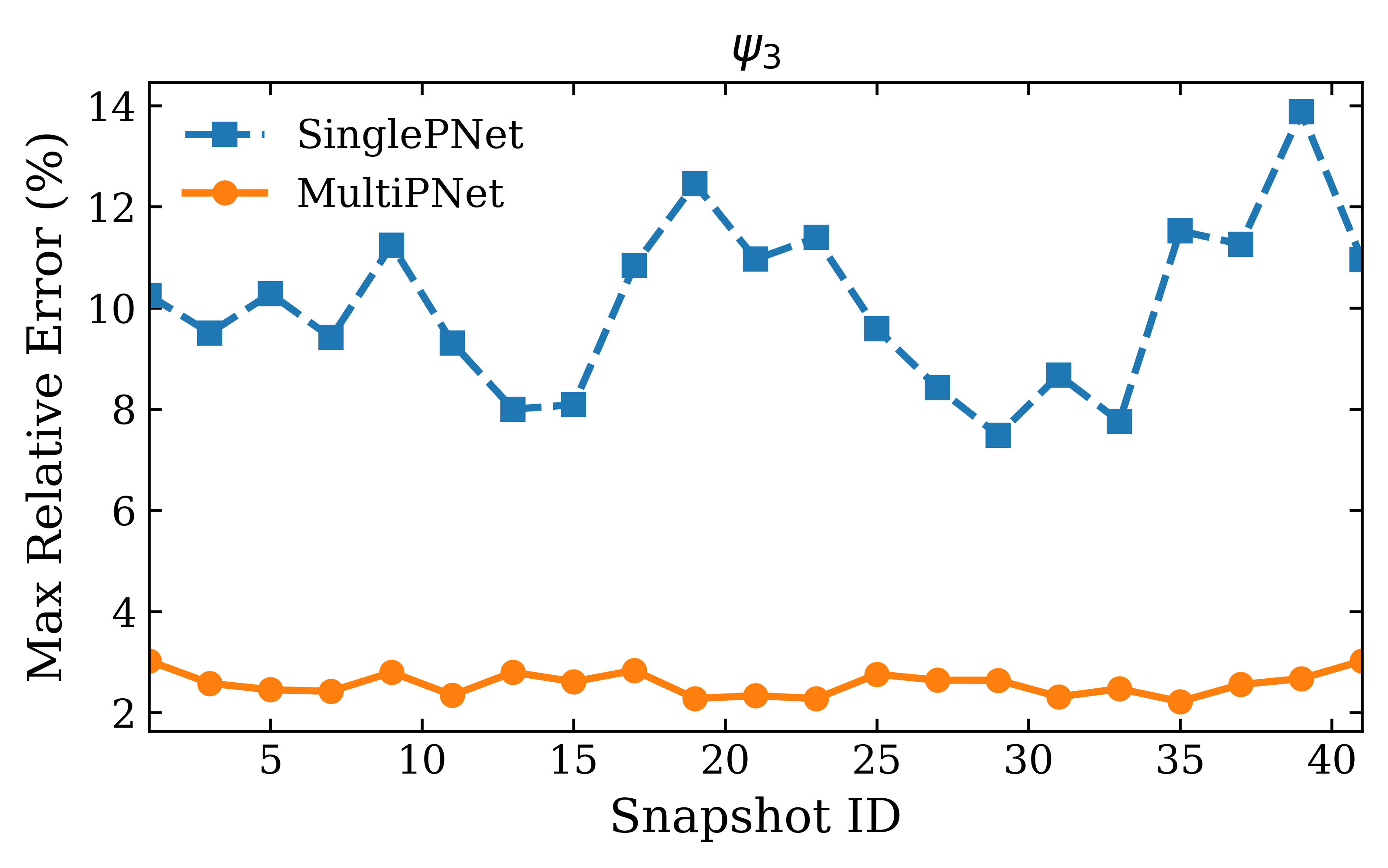}
\end{subfigure}

\vspace{0.5em}

\begin{subfigure}{0.48\textwidth}
    \centering
    \includegraphics[width=\linewidth]{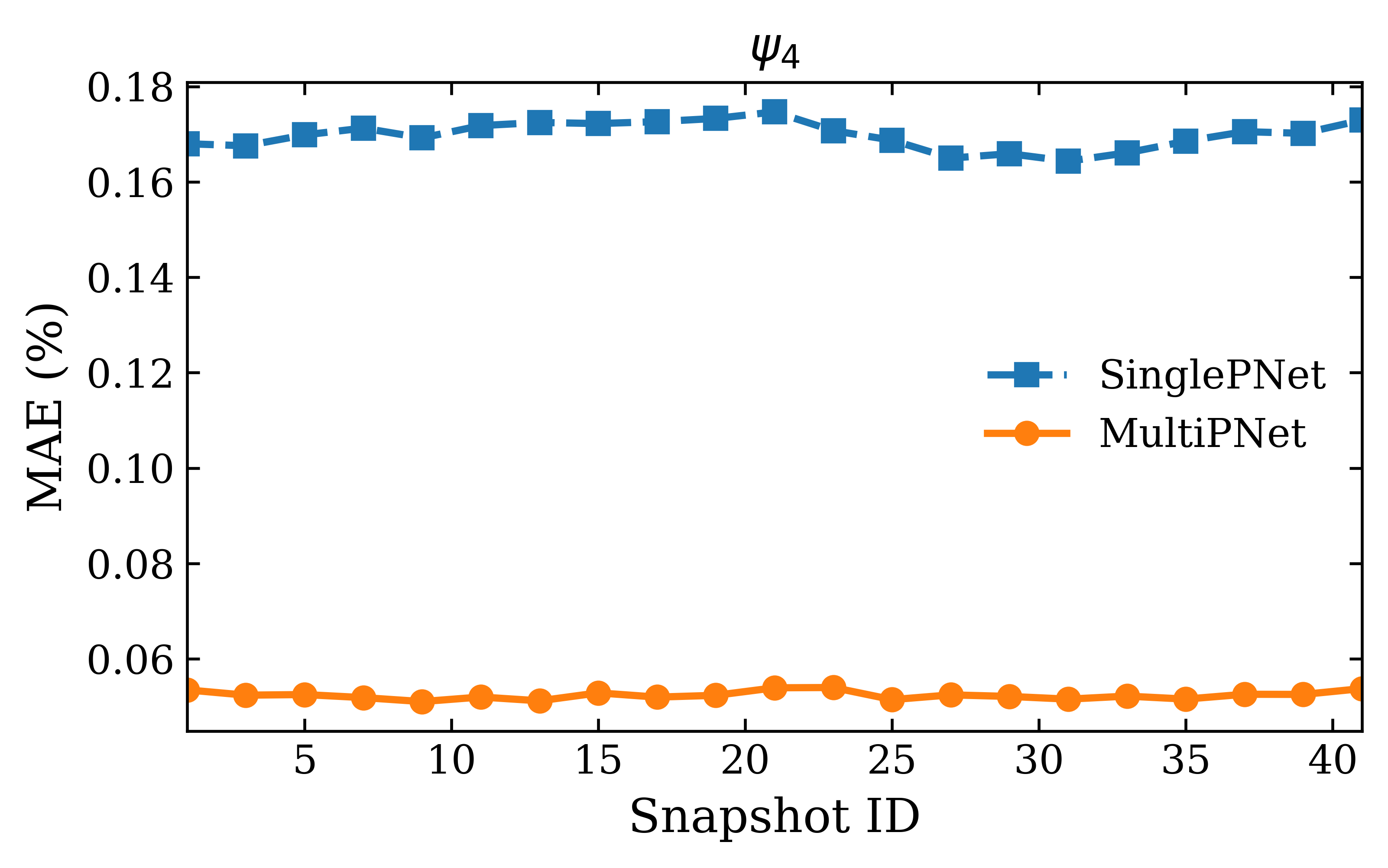}
\end{subfigure}
\hfill
\begin{subfigure}{0.48\textwidth}
    \centering
    \includegraphics[width=\linewidth]{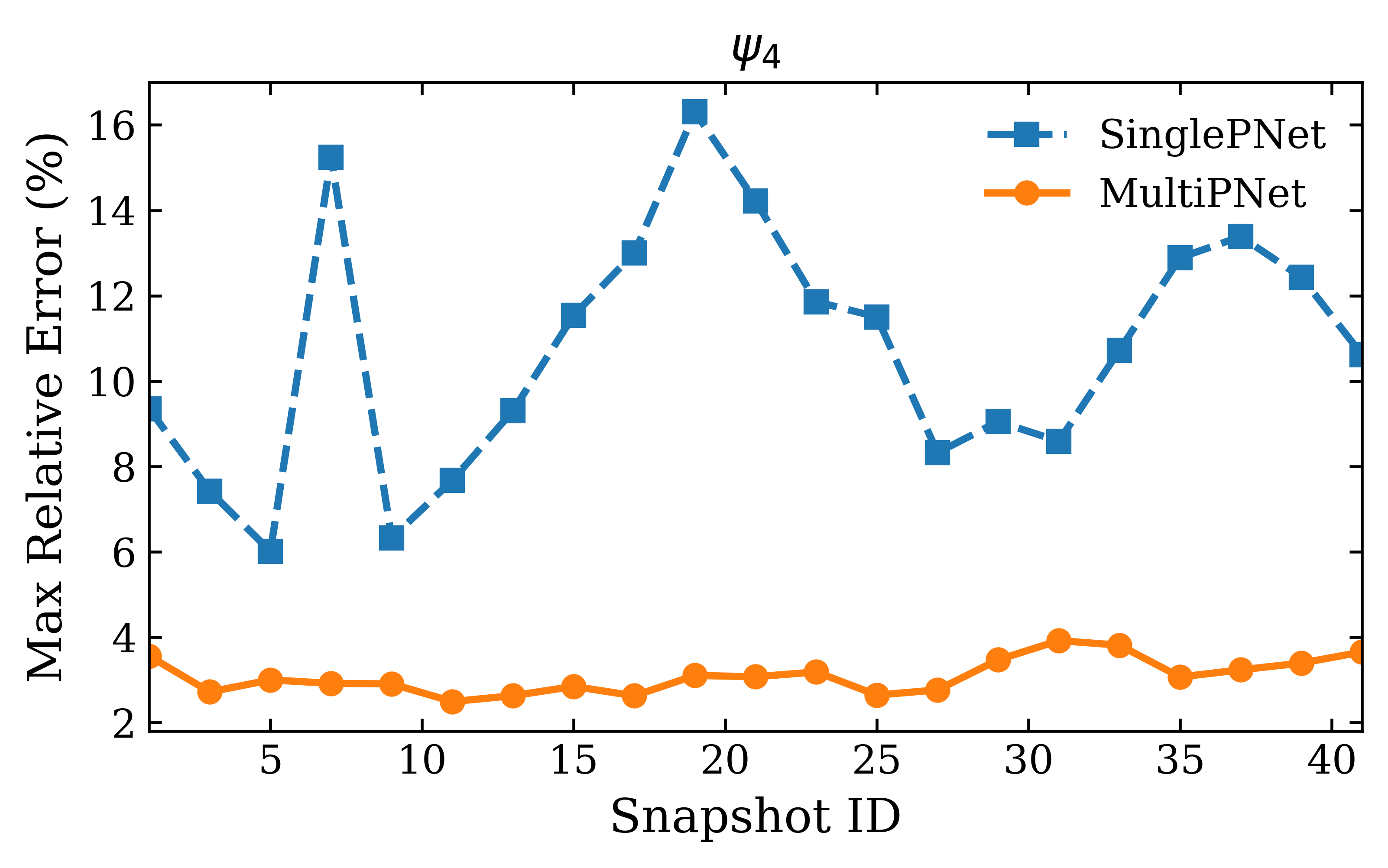}
\end{subfigure}

\caption{Super-resolution error comparison between the original NIF (SinglePNet) and the proposed NIF framework (MultiPNet) on $\psi$'s.}
\label{fig:NIF-error comparison}
\end{figure}

\subsection{Comparison with Single-ParameterNet NIF}\label{subsec:comparison}

In this section, the performance of the proposed multi-ParameterNet NIF framework is compared against the single-ParameterNet NIF~\cite{Pan2023} architecture. For brevity, these models are hereafter referred to as MultiPNet and SinglePNet, respectively. Both models are trained using identical datasets, training procedures, and approximately the same number of trainable parameters (Table~\ref{tab:nif_training}), ensuring that any observed differences in performance are solely attributable to the architectural design.

Figure~\ref{fig:trainingloss} presents the training loss evolution for both models. It is observed that the MultiPNet framework converges significantly faster than the SinglePNet model. For a given loss threshold, the MultiPNet model requires approximately $45\%$ fewer training epochs to reach convergence. This accelerated convergence indicates improved learning efficiency, which can be attributed to the structured latent representation in the proposed framework. By assigning separate ParameterNets to each input component, the model reduces the complexity of learning a joint representation from concatenated inputs, thereby facilitating more effective optimization.

To further assess reconstruction performance, Figure~\ref{fig:NIF-error comparison} compares the MAE and MRE across all nonlinear principal components for both models over the entire dataset. In this figure, the snapshot ID corresponds to the temporal index of the DNS snapshot, where a total of 41 time snapshots are evaluated. The MultiPNet framework consistently achieves lower MAE values across all $\psi$'s and time instances. In addition, the maximum relative errors, computed with respect to the range of each $\psi$, are significantly reduced compared to the SinglePNet model, with improved stability and reduced variability across snapshots.

The superior performance of the proposed framework can be attributed to its ability to learn disentangled latent representations. In the SinglePNet architecture, the coarse-grid inputs are concatenated into a single high-dimensional vector, which forces the model to learn a coupled latent representation. This often leads to latent entanglement, where distinct features from different input components interfere with each other, reducing the model’s ability to accurately capture individual contributions. In contrast, the MultiPNet framework assigns a dedicated ParameterNet to each input component, allowing each latent representation to specialize in capturing its corresponding features. The subsequent fusion of these latent representations enables the model to combine information in a structured and efficient manner.

As a result, the MultiPNet framework not only improves reconstruction accuracy but also enhances training efficiency and model robustness. These findings demonstrate that the structured latent decomposition introduced in the MultiPNet NIF framework provides a more effective representation for multi-component super-resolution tasks compared to the original SinglePNet formulation.

\subsection{Reconstruction of Representative Scalars}\label{subsec:Recon-scalars}

\begin{figure}[tbp]
    \centering
    \begin{subfigure}{\linewidth}
        \centering
        \includegraphics[width=0.8\linewidth]{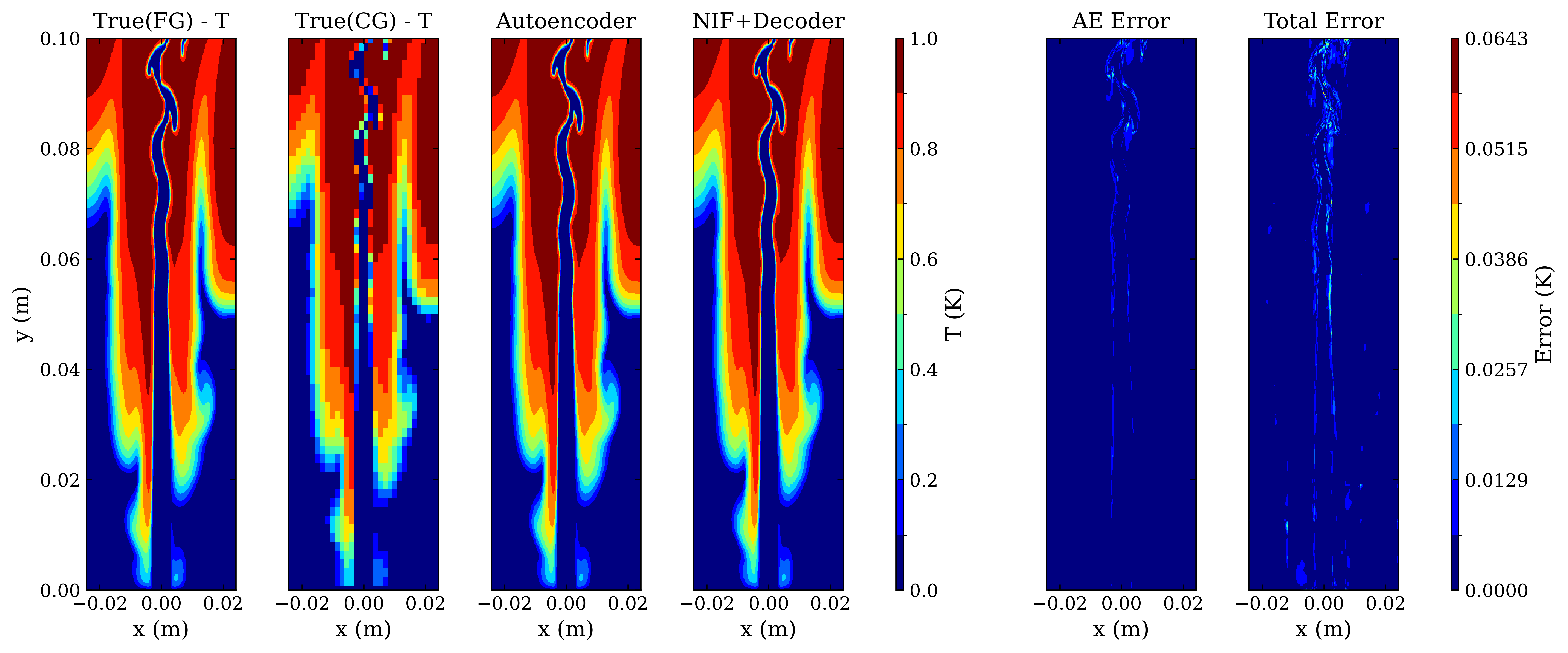}
    \end{subfigure}
    \begin{subfigure}{\linewidth}
        \centering
        \includegraphics[width=0.8\linewidth]{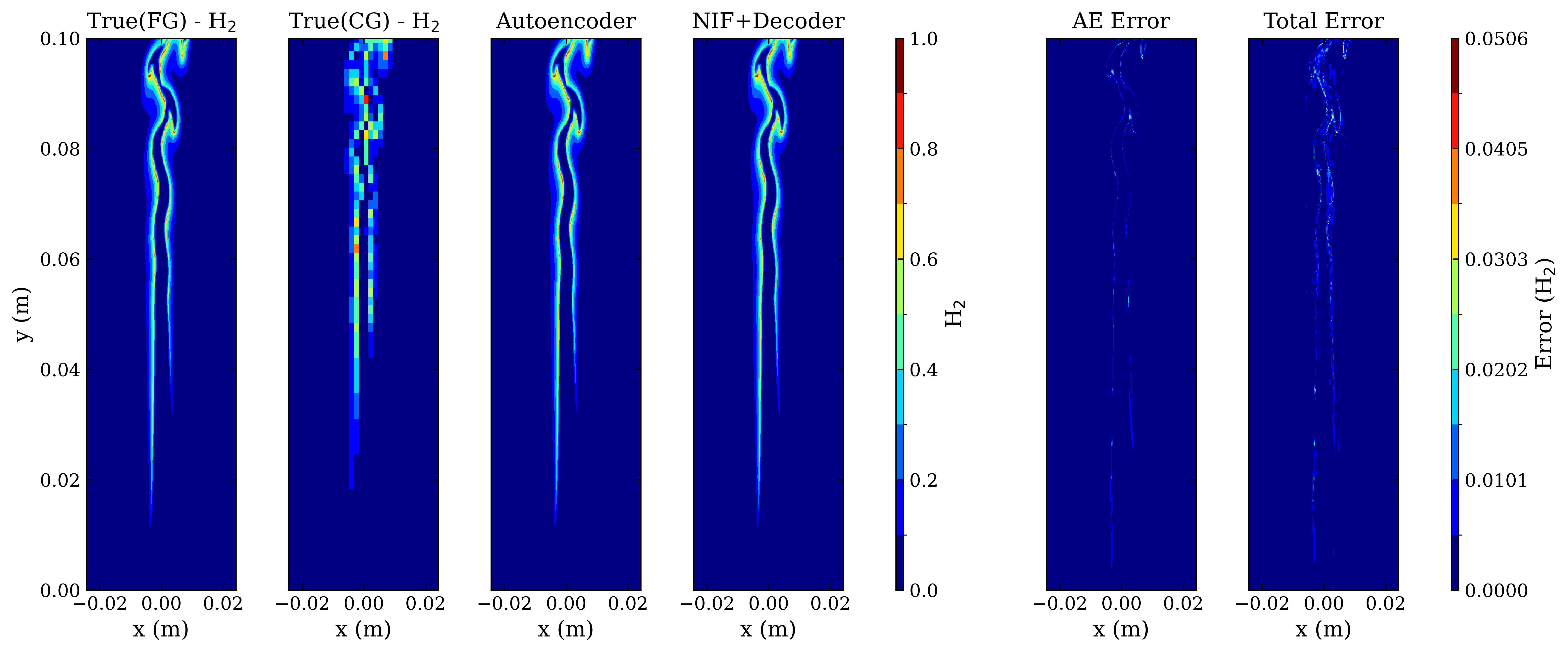}
    \end{subfigure}
    \begin{subfigure}{\linewidth}
        \centering
        \includegraphics[width=0.8\linewidth]{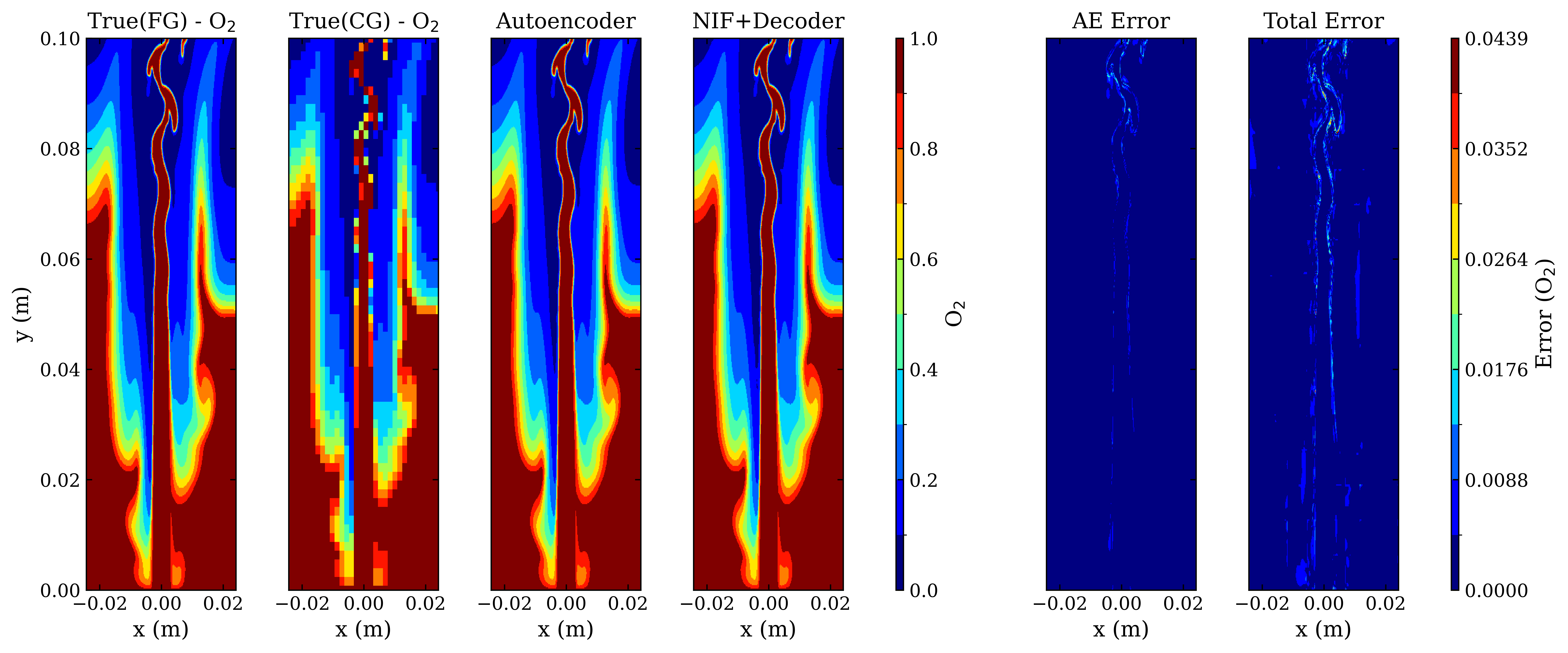}
    \end{subfigure}

    \caption{Reconstruction of normalized representative scalars (temperature, mass fractions of H$_2$ and O$_2$). For each variable, the columns show the true fine-grid (FG) field, coarse-grid (CG) input, autoencoder reconstruction from ground-truth, and final reconstruction obtained by decoding the NIF-predicted FG NLPCs. The last two columns show the corresponding absolute errors for the autoencoder-only and complete NIF--decoder reconstructions.}
    \label{fig:representative-scalar-pred1}
\end{figure}

\begin{figure}[tbp]
    \centering
    \begin{subfigure}{\linewidth}
        \centering
        \includegraphics[width=0.8\linewidth]{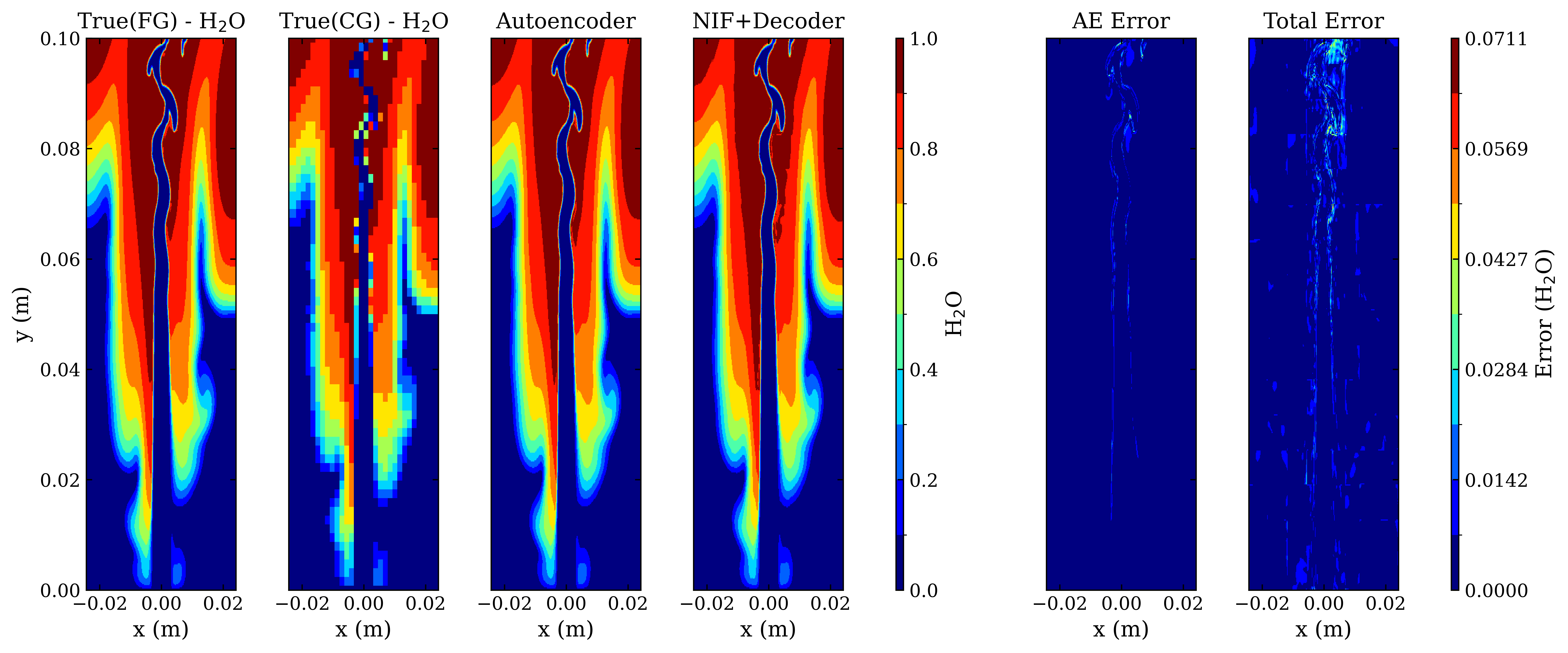}
    \end{subfigure}
    \begin{subfigure}{\linewidth}
        \centering
        \includegraphics[width=0.8\linewidth]{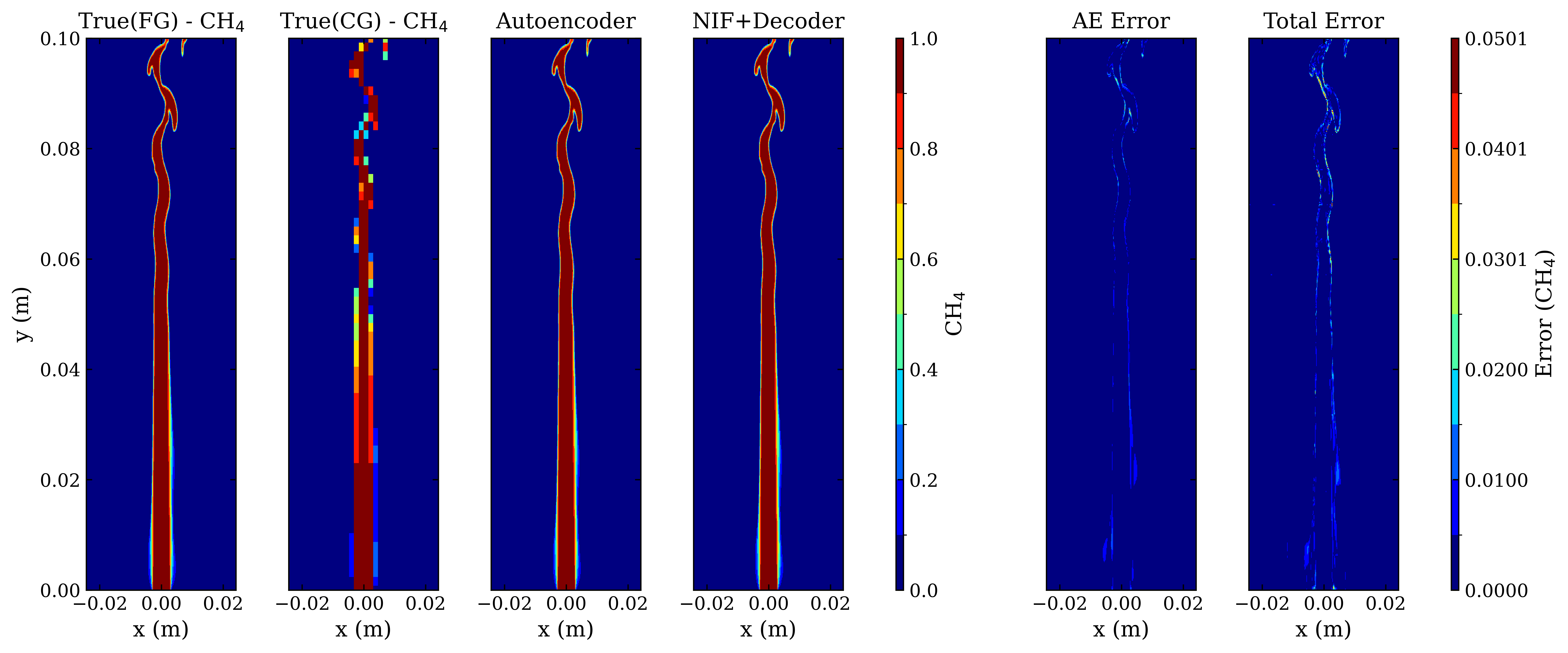}
    \end{subfigure}
    \begin{subfigure}{\linewidth}
        \centering
        \includegraphics[width=0.8\linewidth]{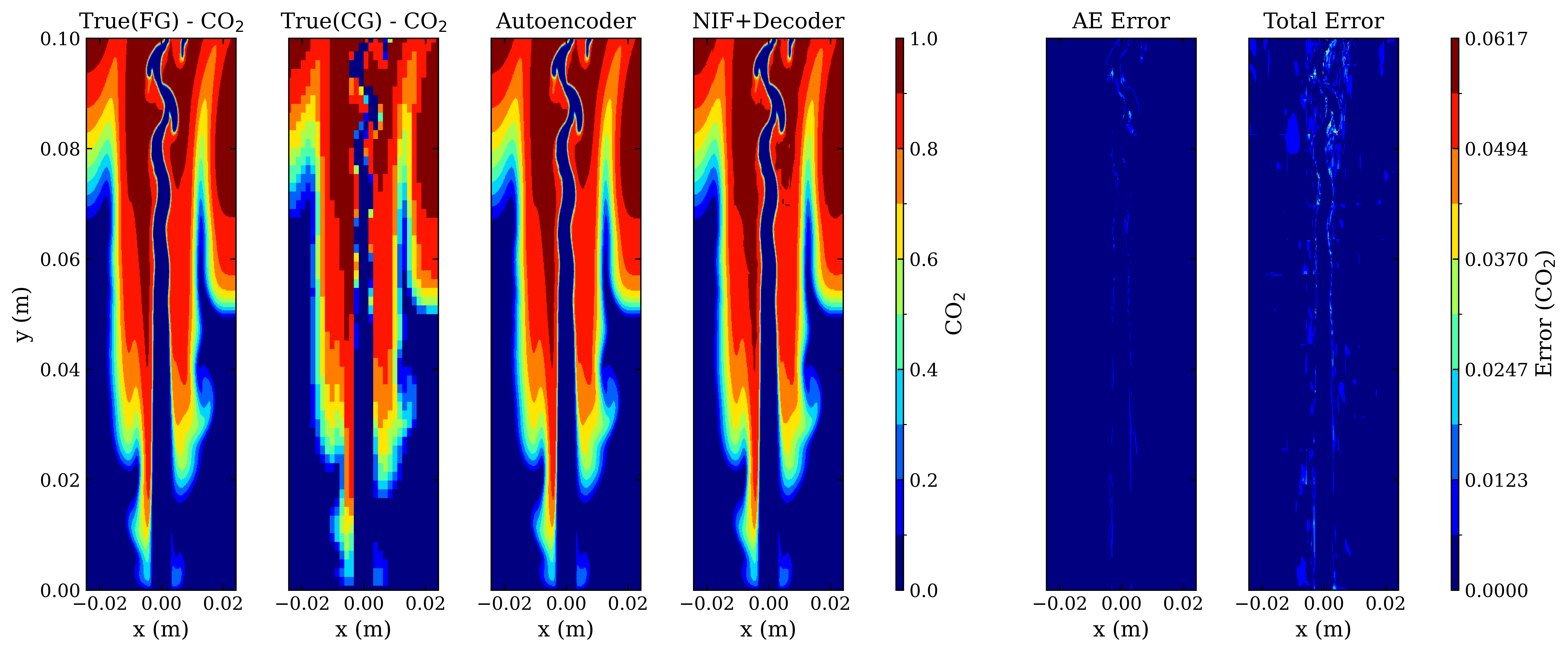}
    \end{subfigure}

    \caption{Reconstruction of normalized representative scalars (H$_2$O, CH$_4$, and CO$_2$). For each variable, the columns show the true fine-grid (FG) field, coarse-grid (CG) input, autoencoder reconstruction from ground-truth, and final reconstruction obtained by decoding the NIF-predicted FG NLPCs. The last two columns show the corresponding absolute errors for the autoencoder-only and complete NIF--decoder reconstructions.}
    \label{fig:representative-scalar-pred2}
\end{figure}

Figures~\ref{fig:representative-scalar-pred1} \& ~\ref{fig:representative-scalar-pred2} present the reconstruction of normalized representative scalars obtained using the proposed framework. Out of the eight representative scalars identified in Section ~\ref{subsec:dataprep}, six are shown here for clarity and visualization purposes, while still providing a comprehensive assessment of reconstruction performance. 

For each scalar, the figures are organized column-wise to enable direct comparison across different stages of the reconstruction process. The first column shows the ground-truth fine-grid (FG) field, representing the reference high-resolution solution. The second column displays the corresponding coarse-grid (CG) field obtained through uniform subsampling with a coarsening factor of 8 in each spatial direction. The third column presents the reconstruction obtained solely from the autoencoder (AE), where the latent nonlinear principal components ($\psi$'s) derived from the ground-truth are decoded. The fourth column shows the final reconstructed field obtained by decoding the $\psi$'s predicted by the proposed NIF super-resolution model.

The last two columns quantify the reconstruction errors. The fifth column corresponds to the autoencoder-only error (AE error), isolating the error introduced by the dimensionality reduction and decoding process. The sixth column represents the total reconstruction error, which includes both the NIF super-resolution error and the autoencoder reconstruction error. This separation enables a clear assessment of the contribution of each component of the framework.

Quantitative metrics further confirm the high accuracy of the proposed framework. The AE-only reconstruction yields very low mean absolute errors (MAE) for all representative scalars (temperature, H$_2$, H, O$_2$, H$_2$O, CH$_4$, CO, and CO$_2$), with values of 0.10\%, 0.05\%, 0.04\%, 0.06\%, 0.12\%, 0.04\%, 0.06\%, and 0.07\%, respectively. The corresponding maximum errors remain below 9.25\% for all scalars. When the NIF super-resolution model is included, the MAE values increase only slightly to 0.22\%, 0.09\%, 0.068\%, 0.16\%, 0.33\%, 0.07\%, 0.18\%, and 0.23\%, while the maximum errors range between 5.92\% and 11.39\%.

To provide a non-learned baseline for the super-resolution task, bicubic interpolation is also applied to the same coarse-grid scalar fields. Bicubic interpolation estimates the fine-grid values using local cubic polynomials in the two spatial directions and therefore provides a smooth reconstruction based only on the available coarse-grid information. Unlike the proposed NIF-based framework, this interpolation procedure does not use any learned representation of the DNS data and does not incorporate correlations among the thermochemical scalars. It therefore provides a useful reference for assessing whether the proposed framework recovers fine-scale scalar structures beyond those obtained through standard interpolation of the coarse-grid fields.

In addition to the pointwise error metrics, the structural fidelity of the reconstructed fields is assessed using the Structural Similarity Index Measure (SSIM), summarized in Table~\ref{tab:ssim_interp_nif}. The bicubic interpolation baseline yields average SSIM values of 0.9143 for both the training and testing datasets, indicating that direct interpolation of the coarse-grid fields can recover the broad spatial organization of the representative scalars. However, the proposed NIF-based reconstruction increases the average SSIM to 0.9929 and 0.9928 for the training and testing datasets, respectively. This substantial improvement demonstrates that the proposed framework is not simply smoothing or interpolating the coarse-grid data, but is recovering additional fine-scale thermochemical structures learned from the DNS data.

For all representative scalars, the NIF-based reconstruction provides consistently higher SSIM values than bicubic interpolation. The improvement is especially evident for scalars with localized reaction-zone structures, such as CH$_4$ and CO. For these variables, the bicubic interpolation SSIM values are approximately 0.894 and 0.899, respectively, while the corresponding NIF reconstructions achieve SSIM values of approximately 0.998 and 0.983. Similar improvements are observed for temperature, H$_2$, H, O$_2$, H$_2$O, and CO$_2$, confirming that the learned reconstruction improves structural fidelity across all scalar components rather than only for a selected subset of variables.

The near-identical NIF SSIM values obtained for the training and testing datasets indicate excellent generalization capability and demonstrate that the model preserves spatial structures with virtually no degradation when evaluated on previously unseen data. The close agreement between the training and testing values, with differences typically on the order of $10^{-4}$, also suggests negligible overfitting. Although H$_2$O and CO exhibit comparatively lower SSIM values than the other scalars, their NIF reconstructions remain well above the bicubic interpolation baseline, indicating that the proposed framework retains strong structural agreement even for variables characterized by steep gradients and localized reaction-zone features.

Despite the increase in MRE introduced by the super-resolution stage, the MAE values remain consistently below 0.35\% across all representative scalars, indicating highly accurate reconstruction throughout the computational domain. The relatively larger maximum errors are associated with a very small number of grid points, typically on the order of 0.01\% of the total domain (approximately 12--15 points out of a $249 \times 505$ grid), and are localized in regions characterized by steep gradients and thin reaction zones.

Even under severe coarsening, corresponding to a reduction factor of 8 in each spatial direction and the associated loss of fine-scale information in the coarse-grid input, the proposed framework successfully reconstructs both the dominant spatial structures and the fine-scale thermochemical variations. More than 99.9\% of the computational domain exhibits reconstruction errors below 2\%, while the consistently high SSIM values demonstrate that the recovered fields retain the spatial organization and morphology of the original fine-grid solutions. Collectively, these results confirm that the proposed NIF-based framework achieves high-fidelity reconstruction of multi-scalar reacting-flow fields while preserving both smooth regions and highly dynamic combustion structures.

\begin{table}
\centering
\small
\caption{Comparison of SSIM values for bicubic interpolation and the proposed NIF reconstruction of representative thermochemical scalars.}
\label{tab:ssim_interp_nif}
\begin{tabular*}{\textwidth}{@{\extracolsep{\fill}}lcccccccc}
\hline
& \multicolumn{4}{c}{\textbf{Training}} &
  \multicolumn{4}{c}{\textbf{Testing}} \\
\cline{2-9}
& \multicolumn{2}{c}{\textbf{Bicubic interp.}} &
  \multicolumn{2}{c}{\textbf{NIF}} &
  \multicolumn{2}{c}{\textbf{Bicubic interp.}} &
  \multicolumn{2}{c}{\textbf{NIF}} \\
\cline{2-9}
\textbf{Scalar} &
\textbf{Mean} & \textbf{Std.} &
\textbf{Mean} & \textbf{Std.} &
\textbf{Mean} & \textbf{Std.} &
\textbf{Mean} & \textbf{Std.} \\
\hline
Temperature & 0.9248 & $6 \times 10^{-4}$ & 0.9933 & $2 \times 10^{-4}$ & 0.9248 & $6 \times 10^{-4}$ & 0.9932 & $3 \times 10^{-4}$ \\
H$_2$       & 0.9328 & $6 \times 10^{-4}$ & 0.9975 & $1 \times 10^{-4}$ & 0.9327 & $5 \times 10^{-4}$ & 0.9974 & $1 \times 10^{-4}$ \\
H           & 0.9129 & $7 \times 10^{-4}$ & 0.9982 & $1 \times 10^{-4}$ & 0.9129 & $7 \times 10^{-4}$ & 0.9982 & $1 \times 10^{-4}$ \\
O$_2$       & 0.9178 & $7 \times 10^{-4}$ & 0.9993 & $1 \times 10^{-4}$ & 0.9178 & $8 \times 10^{-4}$ & 0.9993 & $1 \times 10^{-4}$ \\
H$_2$O      & 0.9145 & $7 \times 10^{-4}$ & 0.9771 & $8 \times 10^{-4}$ & 0.9144 & $7 \times 10^{-4}$ & 0.9768 & $7 \times 10^{-4}$ \\
CH$_4$      & 0.8941 & $9 \times 10^{-4}$ & 0.9978 & $2 \times 10^{-4}$ & 0.8941 & $9 \times 10^{-4}$ & 0.9978 & $2 \times 10^{-4}$ \\
CO          & 0.8988 & $9 \times 10^{-4}$ & 0.9834 & $4 \times 10^{-4}$ & 0.8988 & $8 \times 10^{-4}$ & 0.9832 & $5 \times 10^{-4}$ \\
CO$_2$      & 0.9186 & $7 \times 10^{-4}$ & 0.9962 & $2 \times 10^{-4}$ & 0.9185 & $7 \times 10^{-4}$ & 0.9961 & $2 \times 10^{-4}$ \\
\hline
\textbf{Average SSIM} &
\multicolumn{2}{c}{\textbf{0.9143}} &
\multicolumn{2}{c}{\textbf{0.9929}} &
\multicolumn{2}{c}{\textbf{0.9143}} &
\multicolumn{2}{c}{\textbf{0.9928}} \\
\hline
\end{tabular*}
\end{table}

\begin{figure}[tbp]
    \centering
    \begin{subfigure}{\linewidth}
        \centering
        \includegraphics[width=0.8\linewidth]{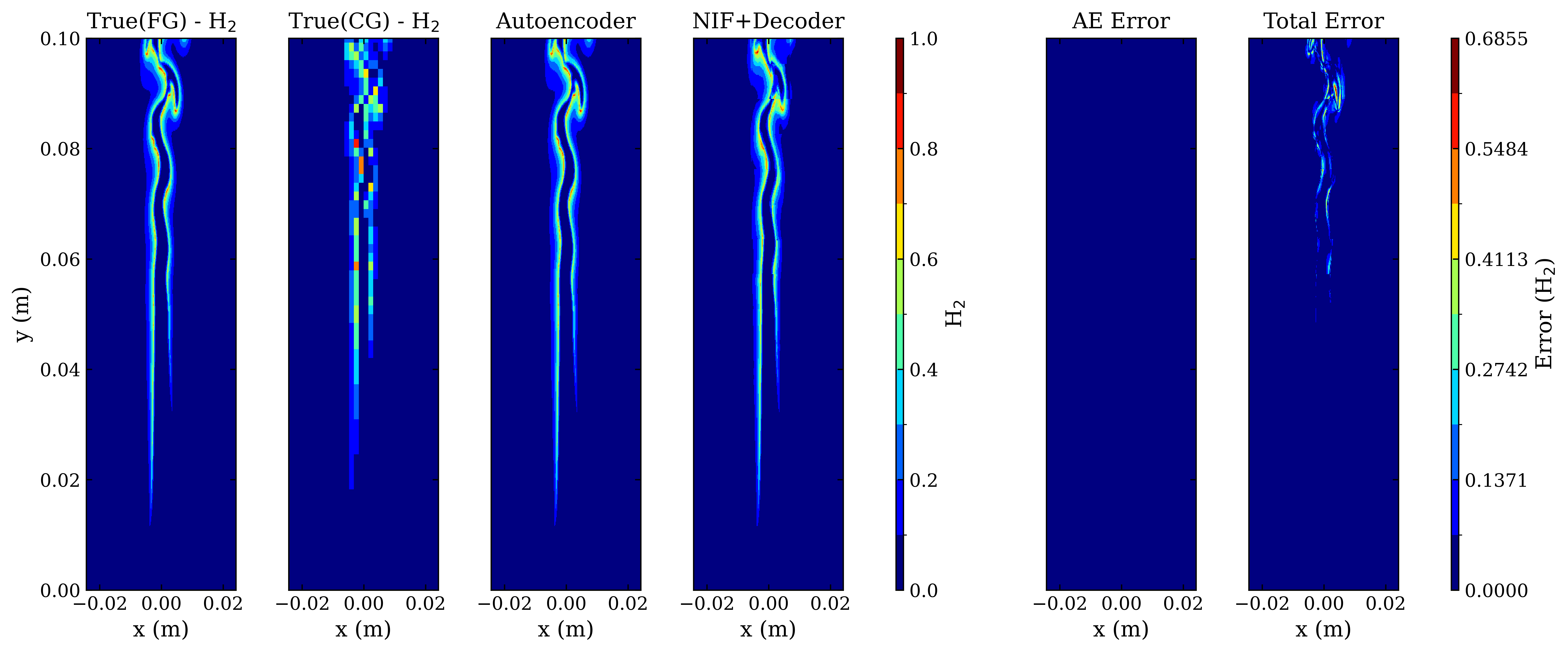}
    \end{subfigure}
    \begin{subfigure}{\linewidth}
        \centering
        \includegraphics[width=0.8\linewidth]{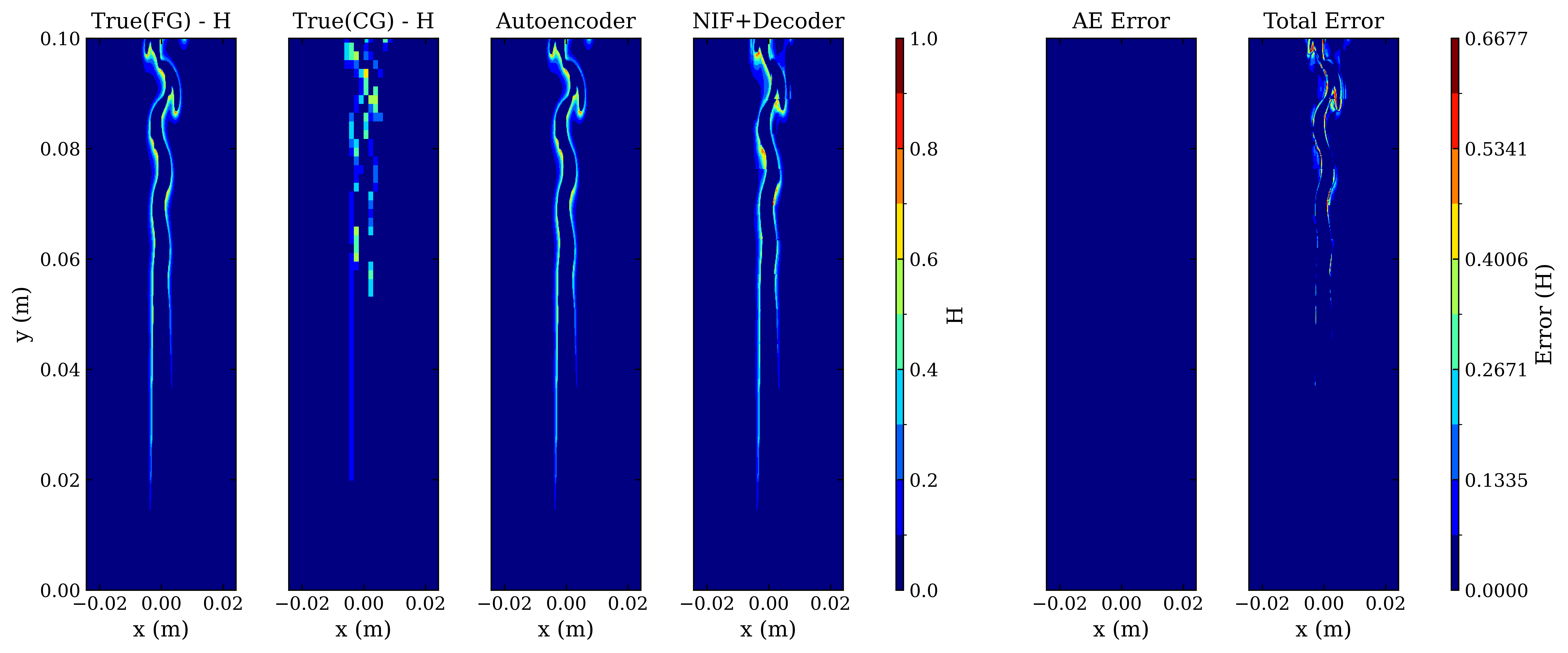}
    \end{subfigure}

    \caption{Representative reconstruction of normalized H$_2$ and H fields for a sample snapshot selected from the additional temporal interval, $t=30.5$--$30.7~\mathrm{ms}$. For each variable, the columns show the true fine-grid (FG) field, coarse-grid (CG) input, autoencoder reconstruction from ground-truth, and final reconstruction obtained by decoding the NIF-predicted FG NLPCs. The last two columns show the corresponding absolute errors for the autoencoder-only and complete NIF--decoder reconstructions.}
    \label{fig:pred-additionaltime}
\end{figure}

To further assess the robustness of the trained model beyond the time interval used in the main training and testing analysis, an additional evaluation is performed over a later time interval, $t = 30.5$--$30.7~\mathrm{ms}$, consisting of 21 snapshots. This assessment is not intended to evaluate temporal forecasting capability, since the proposed framework does not advance the solution in time. Instead, the model is provided with the coarse-grid scalar fields at each later snapshot and is evaluated only as a spatial super-resolution model. Therefore, this test examines whether the learned coarse-to-fine mapping remains effective when similar thermochemical structures appear at a later stage of the flame evolution.

Figure~\ref{fig:pred-additionaltime} shows representative reconstructions of H$_2$ and H from this additional time interval. These two scalars are selected because they are strongly localized in the reaction zone and therefore provide a stringent visual assessment of whether the reconstructed fields preserve thin, highly localized combustion structures. Compared with the main testing interval, the later-time fields exhibit changes in morphology, including more elongated structures and localized regions where some features become weaker or displaced. As a result, the pointwise maximum errors increase, especially near steep gradients. However, the reconstructed fields still capture the dominant reaction-zone topology and recover substantially more spatial structure than the coarse-grid input.

The quantitative results for this additional interval are summarized in Table~\ref{tab:temporal_shift_ssim}. The complete NIF--decoder reconstruction yields MAE values below $1.35\%$ for all representative scalars, with an overall MAE of $0.92\%$. The average SSIM of the NIF-based reconstruction is 0.9558, compared with 0.9150 for bicubic interpolation. Although these SSIM values are lower than those obtained for the main testing interval, the NIF-based reconstruction remains consistently more accurate than bicubic interpolation for all scalars. This indicates that the model retains useful generalization capability for later snapshots containing similar thermochemical structures, while also highlighting that the present framework should be interpreted as a spatial super-resolution model rather than a temporal extrapolation model.

The additional temporal-interval assessment also suggests a practical pathway for extending the framework to more diverse turbulent combustion configurations. Since turbulent reacting flows can produce a wide range of flame structures depending on operating conditions, inlet forcing, turbulence intensity, and geometry, a model trained on one dataset may not fully represent all possible local thermochemical features. However, the ability of the present model to retain useful reconstruction accuracy for later snapshots indicates that the learned coarse-to-fine mapping can provide a suitable initialization for transfer learning. In future applications, the trained NIF framework could therefore be fine-tuned on additional datasets containing new flame structures, rather than trained from scratch, potentially reducing the amount of training data and computational cost required for adaptation to new reacting-flow conditions.

\begin{table}
\centering
\small
\caption{Additional temporal-interval assessment over $t = 30.5$--$30.7~\mathrm{ms}$. The MAE is reported for the proposed NIF framework, and the SSIM values compare bicubic interpolation with the proposed NIF framework.}

\label{tab:temporal_shift_ssim}
\begin{tabular*}{0.9\textwidth}{@{\extracolsep{\fill}}lccccc}
\hline
& \textbf{NIF MAE} &
\multicolumn{2}{c}{\textbf{Bicubic interp. SSIM}} &
\multicolumn{2}{c}{\textbf{NIF SSIM}} \\
\cline{3-6}
\textbf{Scalar} &
\textbf{(\%)} &
\textbf{Mean} & \textbf{Std.} &
\textbf{Mean} & \textbf{Std.} \\
\hline
Temperature & 0.98 & 0.9239 & $2 \times 10^{-4}$ & 0.9589 & $3.1 \times 10^{-3}$ \\
H$_2$       & 0.69 & 0.9334 & $4 \times 10^{-4}$ & 0.9685 & $4.1 \times 10^{-3}$ \\
H           & 0.67 & 0.9139 & $8 \times 10^{-4}$ & 0.9612 & $4.3 \times 10^{-3}$ \\
O$_2$       & 0.93 & 0.9186 & $4 \times 10^{-4}$ & 0.9656 & $2.9 \times 10^{-3}$ \\
H$_2$O      & 1.33 & 0.9141 & $3 \times 10^{-4}$ & 0.9276 & $5.8 \times 10^{-3}$ \\
CH$_4$      & 0.69 & 0.8952 & $4 \times 10^{-4}$ & 0.9689 & $2.3 \times 10^{-3}$ \\
CO          & 0.97 & 0.9011 & $6 \times 10^{-4}$ & 0.9422 & $5.9 \times 10^{-3}$ \\
CO$_2$      & 1.14 & 0.9199 & $3 \times 10^{-4}$ & 0.9531 & $3.7 \times 10^{-3}$ \\
\hline
\textbf{Average} &
\textbf{0.92} &
\multicolumn{2}{c}{\textbf{0.9150}} &
\multicolumn{2}{c}{\textbf{0.9558}} \\
\hline
\end{tabular*}
\end{table}
\section{Conclusions}
A framework for super-resolution of thermochemical scalar fields based on an extension of NIF is proposed. The framework predicts the super-resolution of a small set of NLPCs, which are used to reconstruct fields of thermochemical scalars, including temperature and species mass fractions. The framework is constructed with multiple ParameterNets to accommodate input from multiple NLPCs and calibrate their contribution in evaluating the performance of the NIF networks. A localized training procedure is adopted to extract the features of the scalar fields, which can be reproduced at different parts of the global domain and under different inlet and flow conditions.

The framework was validated in 2D using unsteady lean premixed methane-air flames in a Bunsen burner configuration. It reconstructs fine-grid contours of NLPCs and thermochemical scalars with a coarsening factor of 8 in a given direction. The super-resolution capability of the present extended NIF architecture can be exploited in different contexts to enrich solution methods based on a coarse grid. Such methods include LES or coarse-grid solutions, such as the ones proposed by Brenner and co-workers~\cite{BarSinai2019, Kochkov2021, Zhuang2021}. Moreover, in combustion DNS, quasi-DNS methods ~\cite{Zirwes2019, Zirwes2021, Li2020, Gadalla2024} based on adopting larger grids than what is normally required for DNS can potentially integrate the NIF-based super-resolution and a procedure for correcting the discretization error in the right-hand side of the velocity and thermochemical scalars' governing equations. The same procedure can also be implemented within the context of embedded DNS~\cite{Domingo2017}, where fine-grid solutions are embedded in coarse-grid solutions, as a tool for initializing fine-grid fields. Solution embedding provides mechanisms to resolve essential phenomena that cannot be resolved in a coarse-grid approach or to construct closure statistics on the fly to feed to the coarse-grained solutions. As a future extension of the present work, we plan to integrate the present framework within an {\it {a posteriori}} solution procedure for coarse-grid simulations or embedded DNS.
%\begin{acknowledgments}
\section*{Acknowledgements}
This work was supported by Aramco Americas under Contract No. CW63247.
%\end{acknowledgments}

\bibliographystyle{unsrtnat}
\bibliography{references}   

@misc{DRM19kazakov_drm,
  author = {Alexander Kazakov and Michael Frenklach},
  title = {Reduced Reaction Mechanism for Methane Combustion (DRM19)},
  howpublished = {\url{https://combustion.berkeley.edu/drm/}},
  year = {1994},
  note = {Accessed: 2026-04-26}
}

@ARTICLE{SSIM,
  author={Zhou Wang and Bovik, A.C. and Sheikh, H.R. and Simoncelli, E.P.},
  journal={IEEE Transactions on Image Processing}, 
  title={Image quality assessment: from error visibility to structural similarity}, 
  year={2004},
  volume={13},
  number={4},
  pages={600-612},
  doi={10.1109/TIP.2003.819861}
  }

@misc{zhou2021,
      title={Image Super-Resolution Quality Assessment: Structural Fidelity Versus Statistical Naturalness}, 
      author={Wei Zhou and Zhou Wang and Zhibo Chen},
      year={2021},
      eprint={2105.07139},
      archivePrefix={arXiv},
      primaryClass={eess.IV},
      url={https://arxiv.org/abs/2105.07139}, 
}

@misc{Blastnet3d,
      title={{Turbulence in focus: Benchmarking scaling behavior of 3D volumetric super-resolution with BLASTNet 2.0 data}}, 
      author={Wai Tong Chung and Bassem Akoush and Pushan Sharma and Alex Tamkin and Ki Sung Jung and Jacqueline H. Chen and Jack Guo and Davy Brouzet and Mohsen Talei and Bruno Savard and Alexei Y. Poludnenko and Matthias Ihme},
      year={2023},
      eprint={2309.13457},
      archivePrefix={arXiv},
      primaryClass={cs.LG},
      url={https://arxiv.org/abs/2309.13457}, 
}

@article{PCA-linear,
title = {The reconstruction of thermo-chemical scalars in combustion from a reduced set of their principal components},
journal = {Combustion and Flame},
volume = {162},
number = {5},
pages = {1650-1652},
year = {2015},
issn = {0010-2180},
doi = {https://doi.org/10.1016/j.combustflame.2014.11.027},
url = {https://www.sciencedirect.com/science/article/pii/S0010218014003812},
author = {Hessam Mirgolbabaei and Tarek Echekki}
}

@article{Kumar2023,
title = {Acceleration of turbulent combustion DNS via principal component transport},
journal = {Combust. Flame},
volume = {255},
pages = {112903},
year = {2023},
author = {Anuj Kumar and Martin Rieth and Opeoluwa Owoyele and Jacqueline H. Chen and Tarek Echekki},
}

@article{Kumar2024,
title = {Combustion chemistry acceleration with DeepONets},
journal = {Fuel},
volume = {365},
pages = {131212},
year = {2024},
author = {Anuj Kumar and Tarek Echekki},
}

@article{ Pan2023,
Author = {Pan, Shaowu and Brunton, Steven L. and Kutz, J. Nathan},
Title = {Neural Implicit Flow: a mesh-agnostic dimensionality reduction paradigm
   of spatio-temp oral data},
Journal = {J. Machine Learn. Res.},
Year = {2023},
Volume = {24},
Pages = {41},
}

@article{Domingo2017,
title = {{DNS and approximate deconvolution as a tool to analyse one-dimensional filtered flame sub-grid scale modelling}},
journal = {Combust. Flame},
volume = {177},
pages = {109-122},
year = {2017},
author = {Pascale Domingo and Luc Vervisch},
}

@article{Zirwes2019,
author = {Zirwes, Thorsten and Zhang, Feichi and Habisreuther, Peter and Bockhorn, Henning and Trimis, Dimosthenis},
title = {Large-Scale Quasi-DNS of Mixed-Mode Turbulent Combustion},
journal = {PAMM},
volume = {19},
pages = {e201900420},
year = {2019}
}

@article{ Zirwes2021,
Author = {Zirwes, Thorsten and Zhang, Feichi and Habisreuther, Peter and
   Hansinger, Maximilian and Bockhorn, Henning and Pfitzner, Michael and
   Trimis, Dimosthenis},
Title = {Identification of Flame Regimes in Partially Premixed Combustion from a
   Quasi-DNS Dataset},
Journal = {Flow Turbul. Combust.},
Year = {2021},
Volume = {106},
Pages = {373-404},
}

@article{Li2020,
title = {A quasi-direct numerical simulation solver for compressible reacting flows},
journal = {Compu.  Fluids},
volume = {213},
pages = {104718},
year = {2020},
author = {Tao Li and Jiaying Pan and Fanfu Kong and Baopeng Xu and Xiaohan Wang},
}

@article{Gadalla2024,
title = {Embedded direct numerical simulation of ignition kernel evolution and flame initiation in dual-fuel spray assisted combustion},
journal = {Combust. Flame},
volume = {259},
pages = {113172},
year = {2024},
author = {Mahmoud Gadalla and Shervin Karimkashi and Islam Kabil and Ossi Kaario and Tianfeng Lu and Ville Vuorinen},
}

@inproceedings{Pele2024,
Author = {de Frahan, Marc T. Henry and Esclapez, Lucas and Rood, Jon and Wimer, Nicholas T. and Mullowney, Paul and Perry, Bruce A. and Owens, Landon and Sitaraman, Hariswaran and Yellapantula, Shashank and Hassanaly, Malik and Rahimi, Mohammad J. and Martin, Michael J. and Doronina, Olga A. and Sreejith, N. A. and Rieth, Martin and Ge, Wenjun and Sankaran, Ramanan and Almgren, Ann S. and Zhang, Weiqun and Bell, John B. and Grout, Ray and Day, Marc S. and Chen, Jacqueline H.},
Editor = {Bader, M and Dubey, A and Lusch, B},
Title = {{The Pele Simulation Suite for Reacting Flows at Exascale}},
Booktitle = {{Proc. 2024 SIAM Conf. Paral. Proc. Sci. Comput., PP}},
Series = {Parallel Processing for Scientific Computing},
Year = {2024},
Pages = {13-25},
Note = {Conference on Parallel Processing for Scientific Computing (PP), Baltimore, MD, MAR 05-08, 2024},
Number = {},
Organization = {SIAM},
}

@article{
BarSinai2019,
author = {Yohai Bar-Sinai  and Stephan Hoyer  and Jason Hickey  and Michael P. Brenner },
title = {Learning data-driven discretizations for partial differential equations},
journal = {Proc. Nat. Acad. Sci.},
volume = {116},
number = {31},
pages = {15344-15349},
year = {2019},
}

@article{Kochkov2021,
author = {Dmitrii Kochkov  and Jamie A. Smith  and Ayya Alieva  and Qing Wang  and Michael P. Brenner  and Stephan Hoyer },
title = {Machine learning–accelerated computational fluid dynamics},
journal = {Proc. Nat. Acad. Sci.},
volume = {118},
number = {21},
pages = {e2101784118},
year = {2021},}

@article{Zhuang2021,
  title = {Learned discretizations for passive scalar advection in a two-dimensional turbulent flow},
  author = {Zhuang, Jiawei and Kochkov, Dmitrii and Bar-Sinai, Yohai and Brenner, Michael P. and Hoyer, Stephan},
  journal = {Phys. Rev. Fluids},
  volume = {6},
  issue = {6},
  pages = {064605},
  year = {2021},
}

@article{Nikolaou2018,
  title = {Scalar flux modeling in turbulent flames using iterative deconvolution},
  author = {Nikolaou, Z. M. and Cant, R. S. and Vervisch, L.},
  journal = {Phys. Rev. Fluids},
  volume = {3},
  pages = {043201},
  year = {2018},
}

@article{Lapeyre2019,
title = {Training convolutional neural networks to estimate turbulent sub-grid scale reaction rates},
journal = {Combust. Flame},
volume = {203},
pages = {255-264},
year = {2019},
author = {Corentin J. Lapeyre and Antony Misdariis and Nicolas Cazard and Denis Veynante and Thierry Poinsot},
}

@article{Seltz2019,
title = {Direct mapping from LES resolved scales to filtered-flame generated manifolds using convolutional neural networks},
journal = {Combust. Flame},
volume = {210},
pages = {71-82},
year = {2019},
author = {Andrea Seltz and Pascale Domingo and Luc Vervisch and Zacharias M. Nikolaou},
}

@inproceedings{Bode2020,
Author = {Bode, Mathis and Gauding, Michael and Kleinheinz, Konstantin and Pitsch,
   Heinz},
Editor = {Weiland, M and Juckeland, G and Alam, S and Jagode, H},
Title = {Deep Learning at Scale for Subgrid Modeling in Turbulent Flows:
   Regression and Reconstruction},
Booktitle = {High Performance Computing: ISC High Performance 2019 International Workshops}, 
Series = {Lecture Notes in Computer Science},
Year = {2020},
Volume = {11887},
Pages = {541-560},
}

@article{Grenga2023,
author = {T. Grenga and L. Nista and C. Schumann and A. N. Karimi and G. Scialabba and A. Attili and H. Pitsch},
title = {Predictive Data-Driven Model Based on Generative Adversarial Network for Premixed Turbulence-Combustion Regimes},
journal = {Combust. Sci. Tech.},
volume = {195},
pages = {3923--3946},
year = {2023},
}

@article{Nista2024,
  title = {Influence of adversarial training on super-resolution turbulence reconstruction},
  author = {Nista, Ludovico and Pitsch, Heinz and Schumann, Christoph D. K. and Bode, Mathis and Grenga, Temistocle and MacArt, Jonathan F. and Attili, Antonio},
  journal = {Phys. Rev. Fluids},
  volume = {9},
  pages = {064601},
  year = {2024},
}

@article{Nista2025,
title = {Parallel implementation and performance of super-resolution generative adversarial network turbulence models for large-eddy simulation},
journal = {Comput. Fluids},
volume = {288},
pages = {106498},
year = {2025},
author = {Ludovico Nista and Christoph D.K. Schumann and Peicho Petkov and Valentin Pavlov and Temistocle Grenga and Jonathan F. MacArt and Antonio Attili and Stoyan Markov and Heinz Pitsch},
}

@article{ LeCun1998,
Author = {LeCun, Y and Bottou, L and Bengio, Y and Haffner, P},
Title = {Gradient-based learning applied to document recognition},
Journal = {Proc. IEEE},
Year = {1998},
Volume = {86},
Pages = {2278-2324},
}

@inproceedings{Wang-ESRGAN,
author = {Wang, Xintao and Yu, Ke and Wu, Shixiang and Gu, Jinjin and Liu, Yihao and Dong, Chao and Qiao, Yu and Loy, Chen Change},
title = {ESRGAN: Enhanced Super-Resolution Generative Adversarial Networks},
year = {2018},
isbn = {978-3-030-11020-8},
publisher = {Springer-Verlag},
address = {Berlin, Heidelberg},
url = {https://doi.org/10.1007/978-3-030-11021-5_5},
booktitle = {Computer Vision – ECCV 2018 Workshops: Munich, Germany, September 8-14, 2018, Proceedings, Part V},
pages = {63–79},
location = {Munich, Germany}
}

@article{Fukami2019, 
title={Super-resolution reconstruction of turbulent flows with machine learning}, 
volume={870}, 
journal={J. Fluid Mech.}, 
author={Fukami, Kai and Fukagata, Koji and Taira, Kunihiko}, 
year={2019}, 
pages={106–120}
}

@article{Goodfellow2020,
author = {Goodfellow, Ian and Pouget-Abadie, Jean and Mirza, Mehdi and Xu, Bing and Warde-Farley, David and Ozair, Sherjil and Courville, Aaron and Bengio, Yoshua},
title = {Generative adversarial networks},
year = {2020},
publisher = {Association for Computing Machinery},
address = {New York, NY, USA},
volume = {63},
journal = {Commun. ACM},
pages = {139–144},
}

@misc{Dash2026,
      title={Super-resolution of turbulent reacting flows on complex meshes using graph neural networks}, 
      author={Priyabrat Dash and Konduri Aditya and Christos E. Frouzakis and Mathis Bode},
      year={2026},
      eprint={2603.01080},
      archivePrefix={arXiv},
      primaryClass={physics.flu-dyn},
      url={https://arxiv.org/abs/2603.01080}, 
}

@article{ Barwey2025,
Author = {Barwey, Shivam and Pal, Pinaki and Patel, Umil and Balin, Riccardo and
   Lusch, Bethany and Vishwanath, Venkatram and Maulik, Romit and
   Balakrishnan, Ramesh},
Title = {Mesh-based super-resolution of fluid flows with multiscale graph neural
   networks},
Journal = {Comput. Meth. Appl. Mech. Engg.},
Year = {2025},
Volume = {443},
Pages = {118072},
}

@article{Amarathunga2026,
title = {{React-NIF: A neural implicit flow-based framework for complex fuel combustion chemistry acceleration}},
journal = {Fuel},
volume = {412},
pages = {138166},
year = {2026},
author = {Dakshaka Vishwajith Amarathunga and Tarek Echekki},
}

\end{document}